\pdfoutput=1

\documentclass[12pt,a4paper]{article}
\usepackage{subfigure}
\usepackage{ifthen} 
\newboolean{pdflatex}
\setboolean{pdflatex}{true} 

\newboolean{articletitles}
\setboolean{articletitles}{true} 

\newboolean{uprightparticles}
\setboolean{uprightparticles}{false} 

\newboolean{inbibliography}
\setboolean{inbibliography}{false} 

\usepackage{microtype}
\usepackage{lineno}  
\usepackage{xspace} 
\usepackage{caption} 

\usepackage{graphicx}  
\usepackage{color}
\usepackage{colortbl}
\graphicspath{{./figs/}{./figs/intervals/}} 

\usepackage{amsmath} 
\usepackage{amssymb}
\usepackage{amsfonts}
\usepackage{upgreek} 

\newcommand*\patchAmsMathEnvironmentForLineno[1]{%
\expandafter\let\csname old#1\expandafter\endcsname\csname #1\endcsname
\expandafter\let\csname oldend#1\expandafter\endcsname\csname
end#1\endcsname
 \renewenvironment{#1}%
   {\linenomath\csname old#1\endcsname}%
   {\csname oldend#1\endcsname\endlinenomath}%
}
\newcommand*\patchBothAmsMathEnvironmentsForLineno[1]{%
  \patchAmsMathEnvironmentForLineno{#1}%
  \patchAmsMathEnvironmentForLineno{#1*}%
}
\AtBeginDocument{%
\patchBothAmsMathEnvironmentsForLineno{equation}%
\patchBothAmsMathEnvironmentsForLineno{align}%
\patchBothAmsMathEnvironmentsForLineno{flalign}%
\patchBothAmsMathEnvironmentsForLineno{alignat}%
\patchBothAmsMathEnvironmentsForLineno{gather}%
\patchBothAmsMathEnvironmentsForLineno{multline}%
}

\def\lhcb {\mbox{LHCb}\xspace}

\def\babar  {\mbox{BaBar}\xspace}

\ifthenelse{\boolean{uprightparticles}}%
{

 \def\Pmu         {\ensuremath{\upmu}\xspace}

 \def\Ppi         {\ensuremath{\uppi}\xspace}

 \def\Ppsi        {\ensuremath{\uppsi}\xspace}

 \def\PDelta      {\ensuremath{\Delta}\xspace}                 
 \def\PXi      {\ensuremath{\Xi}\xspace}                 
 \def\PLambda      {\ensuremath{\Lambda}\xspace}                 
 \def\PSigma      {\ensuremath{\Sigma}\xspace}                 
 \def\POmega      {\ensuremath{\Omega}\xspace}                 
 \def\PUpsilon      {\ensuremath{\Upsilon}\xspace}                 
 
 \def\PB      {\ensuremath{\mathrm{B}}\xspace}                 
                  
 \def\PD      {\ensuremath{\mathrm{D}}\xspace}

 \def\PJ      {\ensuremath{\mathrm{J}}\xspace}                 
 \def\PK      {\ensuremath{\mathrm{K}}\xspace}

 \def\Pb      {\ensuremath{\mathrm{b}}\xspace}                 
 \def\Pc      {\ensuremath{\mathrm{c}}\xspace}

 \def\Pi      {\ensuremath{\mathrm{i}}\xspace}

 \def\Ps      {\ensuremath{\mathrm{s}}\xspace}

}
{

 \def\Pmu         {\ensuremath{\mu}\xspace}

 \def\Ppi         {\ensuremath{\pi}\xspace}

 \def\Ppsi        {\ensuremath{\psi}\xspace}                 
                  
 \mathchardef\PDelta="7101
 \mathchardef\PXi="7104
 \mathchardef\PLambda="7103
 \mathchardef\PSigma="7106
 \mathchardef\POmega="710A
 \mathchardef\PUpsilon="7107
                  
 \def\PB      {\ensuremath{B}\xspace}                 
                  
 \def\PD      {\ensuremath{D}\xspace}

 \def\PJ      {\ensuremath{J}\xspace}                 
 \def\PK      {\ensuremath{K}\xspace}

 \def\Pb      {\ensuremath{b}\xspace}                 
 \def\Pc      {\ensuremath{c}\xspace}

 \def\Pi      {\ensuremath{i}\xspace}

 \def\Ps      {\ensuremath{s}\xspace}

}

\def\mup        {\ensuremath{\Pmu^+}\xspace}
\def\mun        {\ensuremath{\Pmu^-}\xspace} 
\def\mumu       {\ensuremath{\Pmu^+\Pmu^-}\xspace}

\def\squark    {\ensuremath{\Ps}\xspace}

\def\cquark    {\ensuremath{\Pc}\xspace}

\def\bquark    {\ensuremath{\Pb}\xspace}

\def\pion  {\ensuremath{\Ppi}\xspace}

\def\pip   {\ensuremath{\pion^+}\xspace}
\def\pim   {\ensuremath{\pion^-}\xspace}

\def\kaon  {\ensuremath{\PK}\xspace}
  \def\Kbar  {\kern 0.2em\overline{\kern -0.2em \PK}{}\xspace}

\def\Kp    {\ensuremath{\kaon^+}\xspace}
\def\Km    {\ensuremath{\kaon^-}\xspace}

\def\KS    {\ensuremath{\kaon^0_{\rm\scriptscriptstyle S}}\xspace} 
 
\def\Kstarz  {\ensuremath{\kaon^{*0}}\xspace}

  \def\Dbar    {\kern 0.2em\overline{\kern -0.2em \PD}{}\xspace}

\def\Dzb     {\ensuremath{\Dbar^0}\xspace}

\def\B       {\ensuremath{\PB}\xspace}
\def\Bbar    {\ensuremath{\kern 0.18em\overline{\kern -0.18em \PB}{}}\xspace}

\def\Bz      {\ensuremath{\B^0}\xspace}
\def\Bzb     {\ensuremath{\Bbar^0}\xspace}
\def\Bu      {\ensuremath{\B^+}\xspace}
\def\Bub     {\ensuremath{\B^-}\xspace}
\def\Bp      {\ensuremath{\Bu}\xspace}
\def\Bm      {\ensuremath{\Bub}\xspace}

\def\Bs      {\ensuremath{\B^0_\squark}\xspace}

\def\jpsi     {\ensuremath{{\PJ\mskip -3mu/\mskip -2mu\Ppsi\mskip 2mu}}\xspace}
\def\psitwos  {\ensuremath{\Ppsi{(2S)}}\xspace}

  \def\Y#1S{\ensuremath{\PUpsilon{(#1S)}}\xspace}

\def\Lz {\ensuremath{\PLambda}\xspace}
\def\Lbar {\ensuremath{\kern 0.1em\overline{\kern -0.1em\PLambda}}\xspace}

\def\Lb      {\ensuremath{\Lz^0_\bquark}\xspace}

\newcommand{\decay}[2]{\ensuremath{#1\!\to #2}\xspace}         

\def\to                 {\ensuremath{\rightarrow}\xspace}

\def\qsq       {\ensuremath{q^2}\xspace}

\def\CP                {\ensuremath{C\!P}\xspace}

\def\AT#1     {\ensuremath{A_{\mathrm{T}}^{#1}}\xspace}           

\def\C#1      {\ensuremath{\mathcal{C}_{#1}}\xspace}                       
\def\Cp#1     {\ensuremath{\mathcal{C}_{#1}^{'}}\xspace}                    
\def\Ceff#1   {\ensuremath{\mathcal{C}_{#1}^{\mathrm{(eff)}}}\xspace}        
\def\Cpeff#1  {\ensuremath{\mathcal{C}_{#1}^{'\mathrm{(eff)}}}\xspace}       
\def\Ope#1    {\ensuremath{\mathcal{O}_{#1}}\xspace}                       
\def\Opep#1   {\ensuremath{\mathcal{O}_{#1}^{'}}\xspace}                    

\newcommand{\tev}{\ifthenelse{\boolean{inbibliography}}{\ensuremath{~T\kern -0.05em eV}\xspace}{\ensuremath{\mathrm{\,Te\kern -0.1em V}}\xspace}}
\newcommand{\gev}{\ensuremath{\mathrm{\,Ge\kern -0.1em V}}\xspace}
\newcommand{\mev}{\ensuremath{\mathrm{\,Me\kern -0.1em V}}\xspace}
\newcommand{\kev}{\ensuremath{\mathrm{\,ke\kern -0.1em V}}\xspace}
\newcommand{\ev}{\ensuremath{\mathrm{\,e\kern -0.1em V}}\xspace}
\newcommand{\gevc}{\ensuremath{{\mathrm{\,Ge\kern -0.1em V\!/}c}}\xspace}
\newcommand{\mevc}{\ensuremath{{\mathrm{\,Me\kern -0.1em V\!/}c}}\xspace}
\newcommand{\gevcc}{\ensuremath{{\mathrm{\,Ge\kern -0.1em V\!/}c^2}}\xspace}
\newcommand{\gevgevcccc}{\ensuremath{{\mathrm{\,Ge\kern -0.1em V^2\!/}c^4}}\xspace}
\newcommand{\mevcc}{\ensuremath{{\mathrm{\,Me\kern -0.1em V\!/}c^2}}\xspace}

\def\mum  {\ensuremath{{\,\upmu\rm m}}\xspace}

\def\invfb   {\ensuremath{\mbox{\,fb}^{-1}}\xspace}

\def\ps   {\ensuremath{{\rm \,ps}}\xspace}

\newcommand{\chisq}{\ensuremath{\chi^2}\xspace}

\def\deriv {\ensuremath{\mathrm{d}}}

\def\gsim{{~\raise.15em\hbox{$>$}\kern-.85em
          \lower.35em\hbox{$\sim$}~}\xspace}
\def\lsim{{~\raise.15em\hbox{$<$}\kern-.85em
          \lower.35em\hbox{$\sim$}~}\xspace}

\def\pt         {\mbox{$p_{\rm T}$}\xspace}

\def\mrad{\ensuremath{\rm \,mrad}\xspace}

\def\evtgen     {\mbox{\textsc{EvtGen}}\xspace}

\def\gauss      {\mbox{\textsc{Gauss}}\xspace}
\def\geant      {\mbox{\textsc{Geant4}}\xspace}

\def\photos     {\mbox{\textsc{Photos}}\xspace}

\def\pythia     {\mbox{\textsc{Pythia}}\xspace}

\def\tell1  {TELL1\xspace}
\def\ukl1   {UKL1\xspace}

\usepackage{hyperref}    
\usepackage[all]{hypcap} 

\usepackage{cite} 
\usepackage{mciteplus}

\begin{document}

\renewcommand{\thefootnote}{\fnsymbol{footnote}}
\setcounter{footnote}{1}


\begin{titlepage}
\pagenumbering{roman}

\vspace*{-1.5cm}
\centerline{\large EUROPEAN ORGANIZATION FOR NUCLEAR RESEARCH (CERN)}
\vspace*{1.5cm}
\hspace*{-0.5cm}
\begin{tabular*}{\linewidth}{lc@{\extracolsep{\fill}}r}
\ifthenelse{\boolean{pdflatex}}
{\vspace*{-2.7cm}\mbox{\!\!\!\includegraphics[width=.14\textwidth]{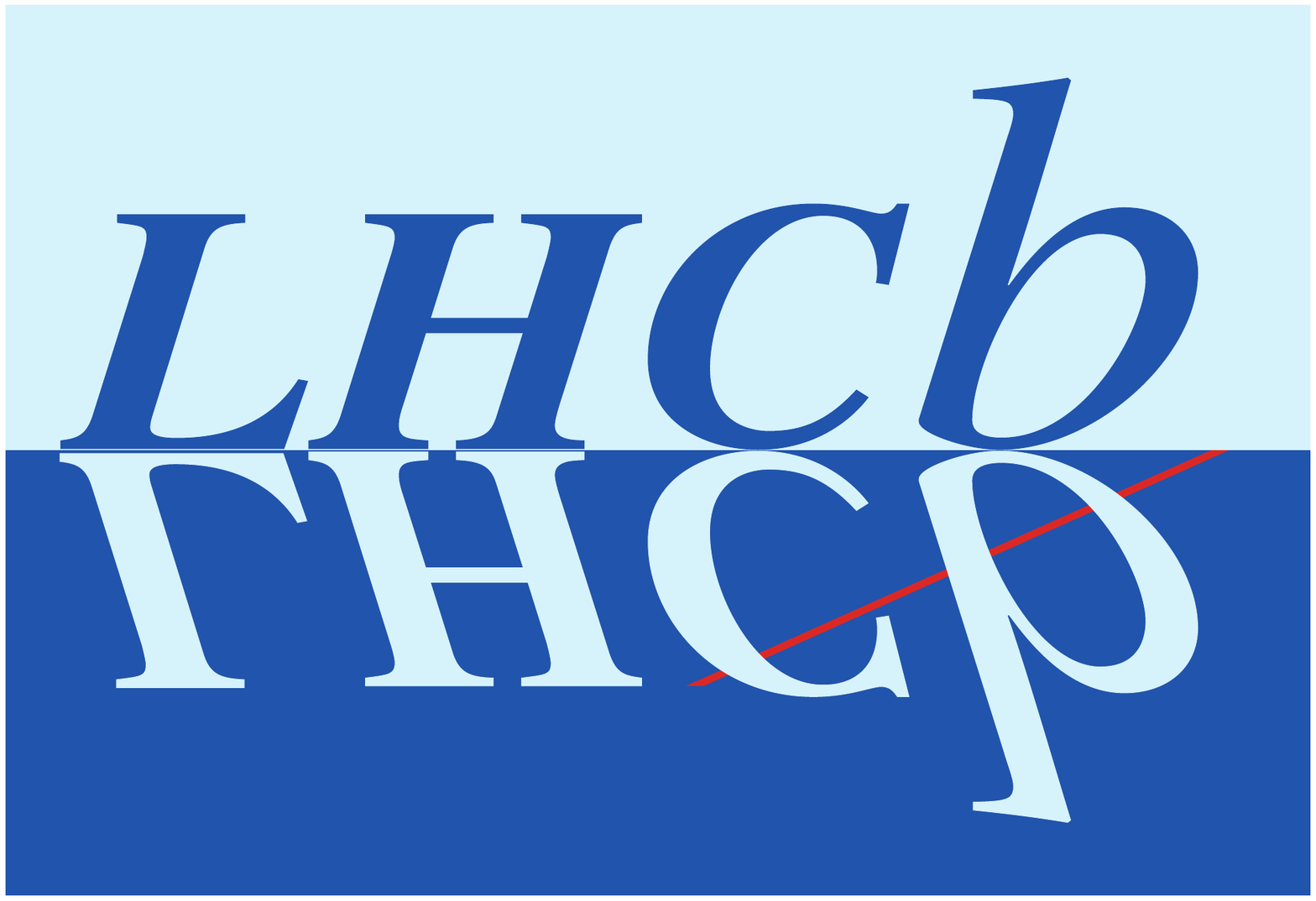}} & &}%
{\vspace*{-1.2cm}\mbox{\!\!\!\includegraphics[width=.12\textwidth]{lhcb-logo.eps}} & &}%
\\
 & & CERN-PH-EP-2014-056 \\  
 & & LHCb-PAPER-2014-007 \\  
 & & 26 May 2014 \\ 
\end{tabular*}

\vspace*{4.0cm}

{\bf\boldmath\huge
\begin{center}
Angular analysis of charged and neutral $B \to K \mu^+\mu^-$ decays
\end{center}
}

\vspace*{2.0cm}

\begin{center}
The LHCb collaboration\footnote{Authors are listed on the following pages.}
\end{center}

\vspace{\fill}

\begin{abstract}
  \noindent
  The angular distributions of the rare decays $B^+ \to K^+\mu^+\mu^-$
  and $B^0 \to K^0_{\rm\scriptscriptstyle S}\mu^+\mu^-$ are studied
  with data corresponding to 3~fb$^{-1}$ of integrated
  luminosity, collected in proton-proton collisions at 7 and 8\tev centre-of-mass energies with the LHCb detector. The angular distribution is described by two
  parameters, $F_{\rm H}$ and the forward-backward asymmetry of the
  dimuon system $A_{\rm FB}$, which are determined in bins of the
   dimuon mass squared. The parameter $F_{\rm H}$ is a measure of the contribution from (pseudo)scalar and tensor amplitudes to the decay width. The measurements of $A_{\rm FB}$ and $F_{\rm H}$ reported here are the most precise to date
  and are compatible with predictions from the Standard Model.
\end{abstract}

\vspace*{2.0cm}

\begin{center}
  Submitted to JHEP 
\end{center}

\vspace{\fill}

{\footnotesize 
\centerline{\copyright~CERN on behalf of the \lhcb collaboration, license \href{http://creativecommons.org/licenses/by/3.0/}{CC-BY-3.0}.}}
\vspace*{2mm}

\end{titlepage}


\newpage
\setcounter{page}{2}
\mbox{~}
\newpage

\centerline{\large\bf LHCb collaboration}
\begin{flushleft}
\small
R.~Aaij$^{41}$, 
B.~Adeva$^{37}$, 
M.~Adinolfi$^{46}$, 
A.~Affolder$^{52}$, 
Z.~Ajaltouni$^{5}$, 
J.~Albrecht$^{9}$, 
F.~Alessio$^{38}$, 
M.~Alexander$^{51}$, 
S.~Ali$^{41}$, 
G.~Alkhazov$^{30}$
% ...

{\footnotesize \it
$ ^{1}$Centro Brasileiro de Pesquisas F\'{i}sicas (CBPF), Rio de Janeiro, Brazil\\
% ...
$ ^{5}$Clermont Universit\'{e}, Universit\'{e} Blaise Pascal, CNRS/IN2P3, LPC, Clermont-Ferrand, France\\
% ...
$ ^{9}$Fakult\"{a}t Physik, Technische Universit\"{a}t Dortmund, Dortmund, Germany\\
% ...
$ ^{30}$Petersburg Nuclear Physics Institute (PNPI), Gatchina, Russia\\
% ...
$ ^{37}$Universidad de Santiago de Compostela, Santiago de Compostela, Spain\\
$ ^{38}$European Organization for Nuclear Research (CERN), Geneva, Switzerland\\
% ...
$ ^{41}$Nikhef National Institute for Subatomic Physics, Amsterdam, The Netherlands\\
% ...
$ ^{46}$H.H. Wills Physics Laboratory, University of Bristol, Bristol, United Kingdom\\
% ...
$ ^{51}$School of Physics and Astronomy, University of Glasgow, Glasgow, United Kingdom\\
$ ^{52}$Oliver Lodge Laboratory, University of Liverpool, Liverpool, United Kingdom\\
% ...
}
\end{flushleft}

\cleardoublepage


\renewcommand{\thefootnote}{\arabic{footnote}}
\setcounter{footnote}{0}



\pagestyle{plain} 
\setcounter{page}{1}
\pagenumbering{arabic}


%

\def\ellell     {\ensuremath{\ell^+ \ell^-}\xspace}


\section{Introduction}
\label{sec:Introduction}

The \decay{\Bp}{\Kp\mumu} and \decay{\Bz}{\KS\mumu} decays are rare,
flavour-changing neutral-current processes that are mediated by
electroweak box and penguin amplitudes in the Standard Model
(SM).\footnote{The inclusion of charge conjugated processes is implied
  throughout.} In well motivated extensions of the SM~\cite{Alok:2008wp,Bobeth:2007dw}, 
  new particles can introduce additional amplitudes that modify the angular distribution of the
   final-state particles predicted by the SM. 

In this paper, the angular distributions of the final-state particles are probed by determining
 the differential rate of the \B meson decays as a function of the angle between the direction
  of one of the muons and the direction of the \Kp or \KS meson in the rest frame of the dimuon
   system. The analysis is performed in bins of \qsq, the dimuon invariant mass squared. 
The angular distribution of \decay{\Bp}{\Kp\mumu} decays has
previously been studied by the \babar~\cite{Aubert:2006vb},
Belle~\cite{:2009zv}, CDF~\cite{Aaltonen:2011ja} and
LHCb~\cite{LHCb-PAPER-2012-024} experiments with less data. 

For the decay \decay{\Bp}{\Kp\mumu}, the differential decay rate can be written as~\cite{Ali:1999mm,Bobeth:2007dw} 

\begin{equation}
  \label{eq:angular:Bp}
  \frac{1}{\Gamma} \frac{\deriv\Gamma}{\deriv\cos\theta_l}
    = \frac{3}{4} (1 - F_{\rm H}) (1 - \cos^2\theta_l) 
    + \frac{1}{2} F_{\rm H} + A_{\rm FB} \cos\theta_l ~,
\end{equation} 

\noindent where $\theta_l$ is the angle between the direction of the \mun (\mup) lepton and the \Kp (\Km) meson for the \Bp (\Bm) decay. The differential decay rate depends on two parameters, the forward-backward
asymmetry of the dimuon system, $A_{\rm FB}$, and a second parameter
$F_{\rm H}$, which corresponds to the fractional contribution of (pseudo)scalar 
and tensor amplitudes to the decay width in the approximation that muons are massless. 
The decay width, $A_{\rm FB}$ and $F_{\rm H}$ all depend on \qsq. 

The structure of Eq.~\ref{eq:angular:Bp} follows from angular momentum conservation in the decay of a pseudo-scalar \B meson into a pseudo-scalar \kaon meson and a pair of muons.  In contrast to the decay \decay{\Bz}{\Kstarz\mumu}, $A_{\rm FB}$ is zero
up to tiny corrections in the SM. A sizable value of $A_{\rm FB}$ is possible in
models that introduce large (pseudo)scalar- or tensor-like
couplings~\cite{Alok:2008wp,Bobeth:2007dw}. The parameter $F_{\rm H}$ 
is non-zero, but small, in the SM due to the finite muon mass. 
For Eq.~\ref{eq:angular:Bp} to remain positive at all lepton angles, $A_{\rm FB}$ and $F_{\rm H}$ have to
satisfy the constraints $0 \leq F_{\rm H} \leq 3$ and $|A_{\rm FB}| \leq F_{\rm
  H}/2$.

Since the \Bz and \Bzb meson can decay to the same $\KS\mumu$ final state, it is not possible to determine the flavour of the \B meson from the
decay products. Without tagging the flavour of the neutral \B meson at production, it is therefore not possible to unambiguously chose the correct muon to determine $\theta_l$. For this reason, $\theta_l$ is always defined with respect to the \mup for decays to the $\KS\mumu$ final-state. In this situation any visible $A_{\rm FB}$ would indicate that there is either a difference in the number of \Bz and \Bzb mesons produced, \CP violation in the decay or that the $A_{\rm FB}$ of the \Bz and \Bzb decay differ. Any residual asymmetry can be canceled by performing the analysis in terms of $|\!\cos\theta_l|$,

\begin{equation}
  \label{eq:angular:B0}
  \frac{1}{\Gamma} \frac{\deriv\Gamma}{\deriv |\!\cos\theta_{l}|} = 
  \frac{3}{2} ( 1 - F_{\rm H} )( 1 - |\!\cos\theta_{l}|^{2} ) + F_{\rm H} ~,
\end{equation}

\noindent where the constraint $0 \leq F_{\rm H} < 3$ is needed for this expression to remain positive
 at all values of $|\!\cos\theta_{l}|$. This simplification of the angular distribution is used for the \decay{\Bz}{\KS\mumu} decay in this paper.

\section{Data and detector description}
\label{sec:Dataset}
The data used for the analysis correspond to 1\invfb of integrated
luminosity collected by the LHCb experiment in $pp$ collisions at $\sqrt{s} = 7\tev$ in 2011 and 2\invfb of integrated luminosity collected
at $\sqrt{s} = 8\tev$ in 2012. The average number of $pp$ interactions, yielding a charged particle 
in the detector acceptance, per bunch crossing was 1.4 in 2011 and 1.7 in 2012.

The \lhcb detector~\cite{Alves:2008zz} is a single-arm forward
spectrometer covering the \mbox{pseudorapidity} range $2<\eta <5$,
designed for the study of particles containing \bquark or \cquark
quarks. The detector includes a high-precision tracking system
consisting of a silicon-strip vertex detector surrounding the $pp$
interaction region, a large-area silicon-strip detector located
upstream of a dipole magnet with a bending power of about
$4{\rm\,Tm}$, and three stations of silicon-strip detectors and straw
drift tubes~\cite{LHCb-DP-2013-003} placed downstream of the magnet.  The combined
tracking system provides a momentum measurement with relative
uncertainty that varies from 0.4\% at 5\gevc to 0.6\% at 100\gevc, and
impact parameter resolution of 20\mum for tracks with large transverse
momentum. Different types of charged hadrons are distinguished by
information from two ring-imaging Cherenkov
detectors~\cite{LHCb-DP-2012-003}. Photon, electron and hadron
candidates are identified by a calorimeter system consisting of
scintillating-pad and preshower detectors, an electromagnetic
calorimeter and a hadronic calorimeter. Muons are identified by a
system composed of alternating layers of iron and multiwire
proportional chambers~\cite{LHCb-DP-2012-002}.

Samples of simulated \decay{\Bp}{\Kp\mumu} and \decay{\Bz}{\KS\mumu}
decays are used to understand how the detector geometry, the
reconstruction and subsequent event selection bias the angular
distribution of the decays.  In the simulation, $pp$ collisions are
generated using \pythia~\cite{Sjostrand:2006za} with a specific \lhcb
configuration~\cite{LHCb-PROC-2010-056}.  Decays of hadronic particles
are described by \evtgen~\cite{Lange:2001uf}, in which final state
radiation is generated using \photos~\cite{Golonka:2005pn}. The
interaction of the generated particles with the detector and its
response are implemented using the \geant
toolkit~\cite{Allison:2006ve, *Agostinelli:2002hh} as described in
Ref.~\cite{LHCb-PROC-2011-006}.

\section{Selection of signal candidates}
\label{sec:Selection} 

The LHCb trigger system~\cite{LHCb-DP-2012-004} consists of a hardware
stage, based on information from the calorimeter and muon systems,
followed by a software stage, which applies a full event
reconstruction. In the hardware stage of the trigger, candidates are
selected with at least one muon candidate with transverse momentum, $\pt >
1.48\,(1.76)\gevc$ in 2011 (2012). In the second stage of the trigger,
at least one of the final-state particles from the \Bz or \Bp meson decay is
required to have $\pt > 1.0\gevc$ and impact parameter larger than
100\mum with respect to any primary vertex (PV) from the
$pp$ interactions in the event. Tracks from two or more of the final-state particles are required to form a secondary vertex that is
displaced from all of the PVs.

The \KS mesons from the decay \decay{\Bz}{\KS\mumu} are reconstructed through their decay \decay{\KS}{\pip\pim} in two different categories: the first category contains
\KS mesons that decay early enough that the final-state pions are
reconstructed in the vertex detector; and the second contains \KS mesons that
decay later, such that the first track segment that can be reconstructed is in the large-area silicon-strip detector. These categories are referred to as
\emph{long} and \emph{downstream}, respectively. Candidates in the long category have better mass, momentum and vertex resolution.

Reconstructed tracks that leave hits in the LHCb muon system are
positively identified as muons. Two muons of opposite charge are then
combined with either a track (\Kp) or a reconstructed \KS to form a
\Bp or \Bz candidate. The $\pip\pim$ pair from the reconstructed \KS is constrained
 to the known \KS mass when determining the mass of the \Bz candidate. 
 Neural networks, using information from the RICH detectors,
calorimeters and muon system, are used to reject backgrounds where
either a pion is misidentified as the kaon in the \Bp decay or a pion
or kaon are incorrectly identified as one of the muons.

An initial selection is applied to \Bp and \Bz candidates to reduce
the level of the background. The selection criteria
are common to those described in Ref.~\cite{LHCb-PAPER-2013-039}: the
$\mu^\pm$ and the \Kp candidates are required to have $\chi^{2}_{\rm IP} > 9$, where $\chi^{2}_{\rm IP}$ is
defined as the minimum change in $\chi^2$ of the vertex fit to any of
the PVs in the event when the particle is added to that PV; the dimuon
pair vertex fit has $\chisq<9$; the \B candidate
is required to have a vertex fit $\chisq < 8$ per degree of freedom; 
the \B momentum vector is aligned with respect to one of the PVs in the event within 14\mrad, the \B candidate has $\chi^{2}_{\rm IP} < 9$ with respect to that
PV and the vertex fit \chisq of that PV increases by more than 121 when including the \B decay products. In addition, the \KS candidate is
required to have a decay time larger than 2\ps. 

The initial selections are followed by tighter multivariate selections,
based on boosted decision trees~(BDTs)~\cite{Breiman} with the
AdaBoost algorithm\cite{AdaBoost}. The working points for the BDTs are 
chosen to maximise $N_{\rm S}/\sqrt{N_{\rm S}+N_{\rm B}}$, where $N_{\rm S}$ and $N_{\rm B}$ are
 the expected numbers of signal and background candidates within 
 $\pm 50\mevcc$ of the known \Bz or \Bp meson masses, respectively. 
For the \decay{\Bp}{\Kp\mumu} decay, the variables used in the BDT are
identical to those of Ref.~\cite{LHCb-PAPER-2013-039}. In contrast to
that analysis, however, the multivariate selection
is trained using a sample of simulated events to model the
signal and candidates from the data with $\Kp\mumu$ invariant masses in the range $5700 <
m(\Kp\mumu) < 6000\mevcc$ for the background. This background sample
is not used in the subsequent analysis, where the invariant mass of
the candidates is restricted to the range $5170 < m(\Kp\mumu) <
5700\mevcc$.  The multivariate selection has an efficiency of 89\% for
signal and removes 94\% of the background that remains after the
initial selection. For the \decay{\Bz}{\KS\mumu} decay, two independent BDTs are trained
for the long and downstream categories. Samples of
simulated events are used in the signal training and candidates from the data with
masses $5700 < m(\KS\mumu) < 6000\mevcc$ for the background
training. The following information is used in the classifiers: the
\Bz candidate momentum and \pt, its vertex quality ($\chi^2$) and decay time, the \KS candidate \pt, and the angle between the \Bz candidate
momentum and the direction between the PV and the \Bz decay
vertex. For the long category, the \KS candidate $\chi^{2}_{\rm IP}$ is also
included.  The multivariate selection removes 99\% of the
combinatorial background and is 66\% and 48\% efficient for the long
and downstream signal categories.


Combinatorial backgrounds for the \decay{\Bp}{\Kp\mumu} decay, where the $\Kp\!$, \mup and \mun candidates do not all
come from the same $b$-hadron decay, are reduced to a small level by
the multivariate selection. After applying the multivariate selection, the 
signal-to-background ratio in a $\pm 50\mevcc$ range around the known \Bp mass is better than six-to-one. Remaining backgrounds mainly come from $b$-hadron decays that are fully or partially reconstructed in the detector. The \decay{\Bp}{\jpsi\Kp} and \decay{\Bp}{\psitwos\Kp} decays\footnote{Throughout this paper the decays \decay{\Bp}{\jpsi\Kp} and \decay{\Bz}{\jpsi\KS} refer to decays of \Bp and \Bz mesons to $\Kp\mumu$ and $\KS\mumu$ final-states, respectively, through the decay \decay{\jpsi}{\mumu}.} are rejected by
removing the regions of dimuon mass around the charmonium
resonances ($8.0 < \qsq < 11.0\gev^{2}/c^{4}$ and $12.5 < \qsq < 15.0\gev^{2}/c^{4}$).
These decays can also form a background to the
\decay{\Bp}{\Kp\mumu} decay if the kaon is incorrectly identified as a
muon and the muon with the same charge is incorrectly identified as a
kaon. This background is removed by rejecting candidates with a
$\Kp\mun$ pair whose invariant mass (under the \mumu mass hypothesis) is consistent
 with that of the \jpsi or \psitwos meson, if the reconstructed kaon can also be matched to hits in the muon
system.  A narrow range in \qsq from $0.98 < \qsq < 1.10\gev^{2}/c^{4}$ is also
 removed to reject \decay{\Bp}{\phi\Kp} decays, followed by the \decay{\phi}{\mumu} decay. The region $m({\Kp\mumu}) < 5170\mevcc$
is contaminated by partially reconstructed $b$-hadron decays such
as \decay{\Bz}{\Kstarz\mumu} where the pion from the
\decay{\Kstarz}{\Kp\pim} decay is not reconstructed. This region is not used in  the subsequent analysis and dictates the lower bound of the $5170 < m({\Kp\mumu}) < 5700\mevcc$ mass range. Backgrounds from fully hadronic $b$-hadron decays, such as
the decay \decay{\Bp}{\Kp\pip\pim}, are reduced to a negligible level
using stringent muon-identification selection criteria. A further requirement is applied
on the $\Kp\mun$ pair to remove a small
contribution from \decay{\Bp}{\Dzb\pip}
decays with \decay{\Dzb}{\Kp\pim}, where the pions survive the muon-identification requirements. 
Candidates are rejected if the mass of the $\Kp\mun$ pair, computed under the $K^+\pi^-$ hypothesis, 
is in the range $1850 < m(\Kp\pim) < 1880\mevcc$. After the
application of all selection criteria, the background from other \bquark-hadron decays 
is reduced to $\mathcal{O}(0.1\%)$ of the level of the signal. The
total efficiency for reconstructing and selecting the \decay{\Bp}{\Kp\mumu} decay is around 2\%.


Due to the long lifetime of the \KS meson, there are very few \bquark-hadron decays that can be mistakenly identified as
\decay{\Bz}{\KS\mumu} decays. The largest source of fully reconstructed background is the decay \decay{\Lb}{\Lz\mumu}, where
the proton from the \decay{\Lz}{p\pim} decay is incorrectly
identified as a \pip. This background is removed by rejecting \KS
meson candidates if the mass of the $\pip\pim$ pair, under the $p\pim$
 mass hypothesis, is consistent with that of a $\Lz$ baryon within $\pm10\mevcc$ ($\pm15\mevcc$) for long (downstream) candidates. This veto is $95\%$ efficient on
genuine \KS meson decays and removes more than 99\% of $\Lz$
baryons. The total efficiency for reconstructing the
\decay{\Bz}{\KS\mumu} decay is about $0.2\%$, which is a factor of ten lower than for the charged
decay. This is due to a combination of three effects: the long flight
distance of \KS mesons in the detector, the \decay{\KS}{\pip\pim} branching fraction, and the requirement of
having four, rather than three, tracks within the detector
acceptance. After applying the selection procedure, the signal-to-background ratio in a $\pm 50\mevcc$ range around the known \Bz mass is better than three-to-one for the \decay{\Bz}{\KS\mumu} decay. 


After applying the full selection criteria, more than $99\%$ of the selected events contain only 
one \Bp or \Bz candidate. Events containing more than one candidate have all but one candidate removed at random in the subsequent analysis.


\section{Angular acceptance}
\label{sec:Acceptance}

The geometrical acceptance of the \lhcb detector, the trigger and the
event selection can all bias the $\cos\theta_l$ distribution of the
selected candidates.  The angular acceptance is
determined using a sample of simulated signal events.  The acceptance as a
function of $\cos\theta_l$ is parameterised using a fourth-order
polynomial function, fixing the odd-order terms to zero so that the
acceptance is symmetric around zero. Any small asymmetry in the acceptance
 for \B and \Bbar mesons, due to charge asymmetries in the reconstruction, cancels when combining \B and \Bbar meson decays.

\begin{figure}[!tb]
  \centering
  \includegraphics[scale=0.37]{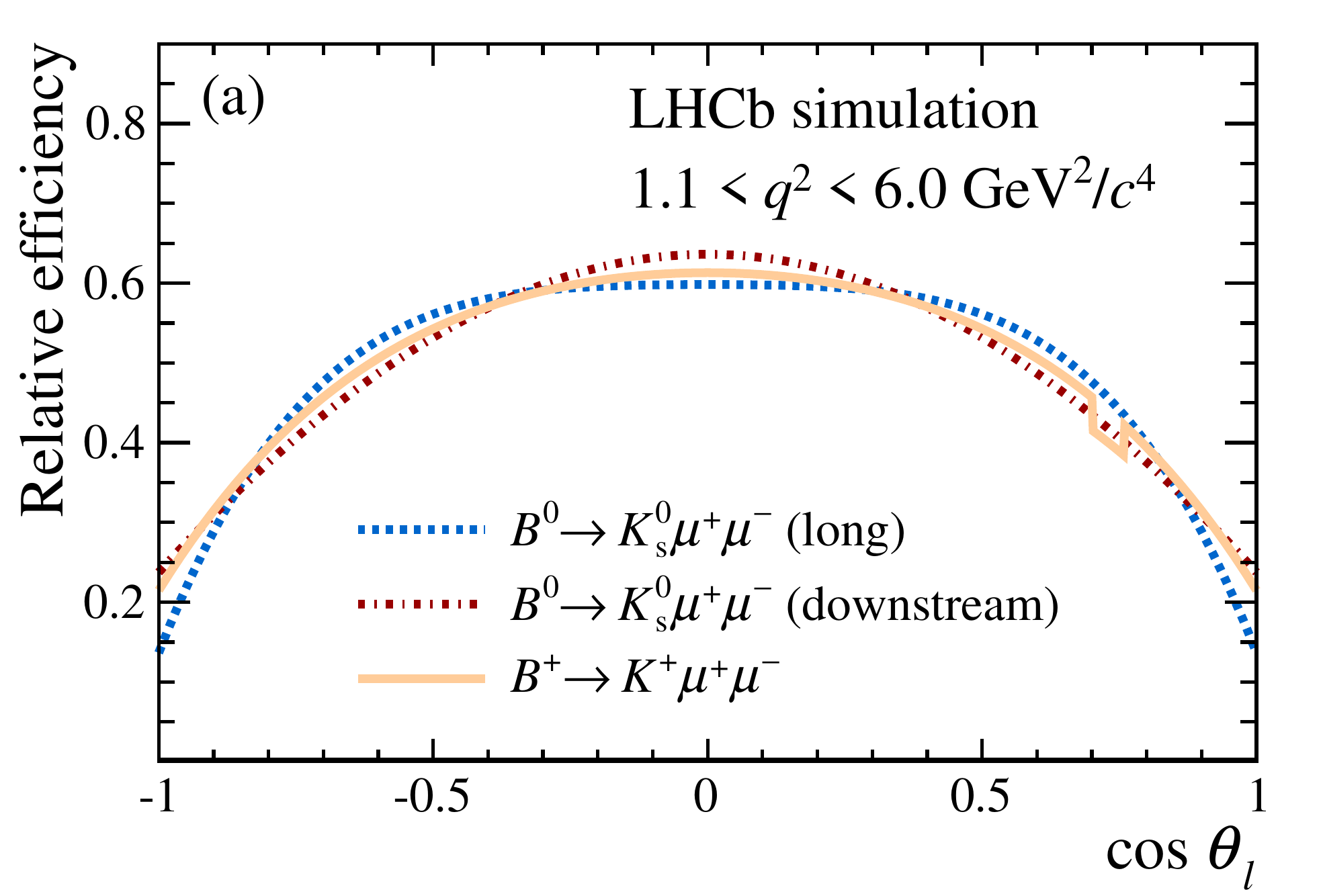}
  \includegraphics[scale=0.37]{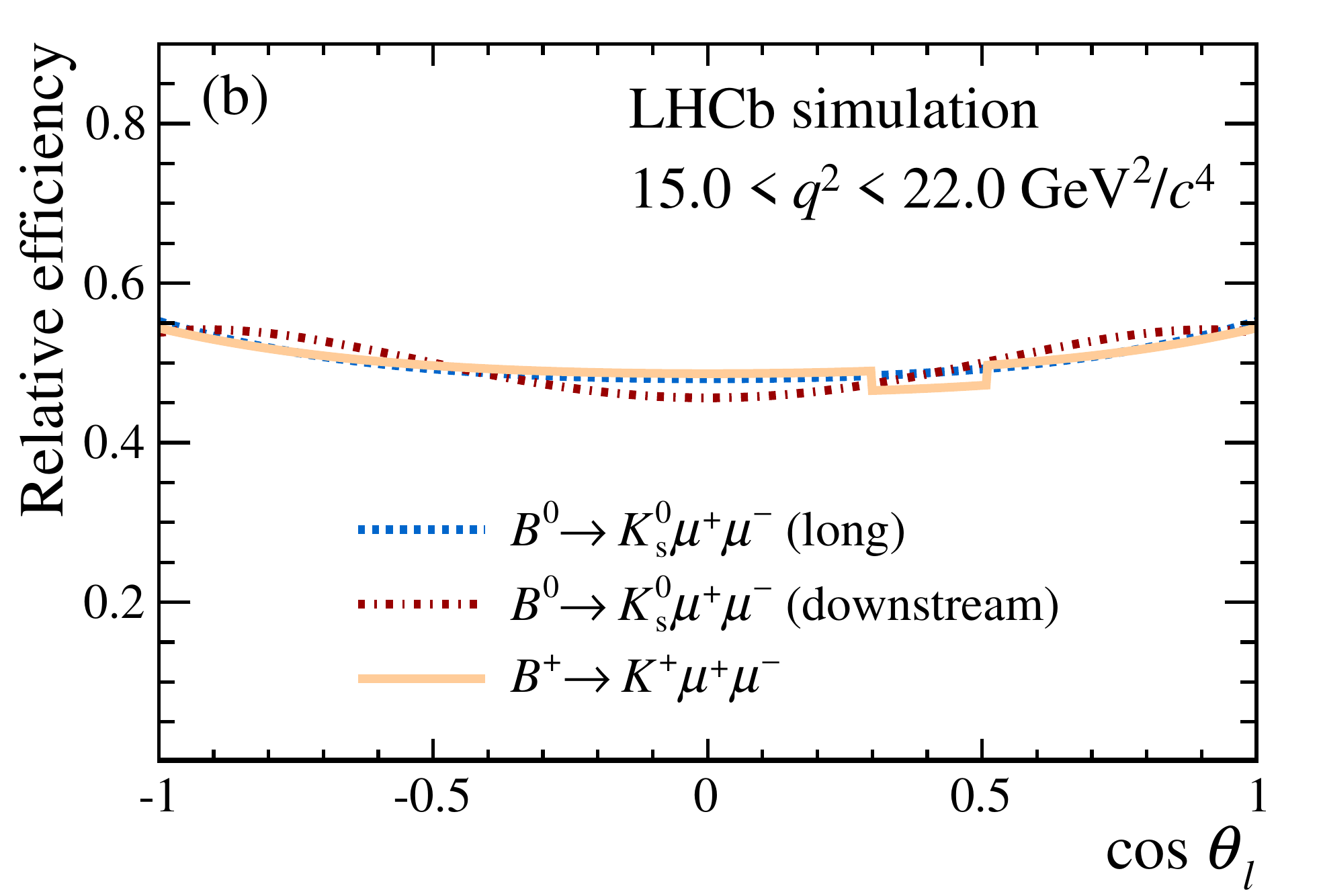}
  \caption{Angular acceptance as derived from simulation in the dimuon mass squared ranges
     (a) $1.1< \qsq < 6.0\gevgevcccc$ and (b) $15.0< \qsq <
    22.0\gevgevcccc$. The dip in the acceptance for \decay{\Bp}{\Kp\mumu} decays results from the veto used to reject \decay{\Bp}{\Dzb\pip} decays (see text). The acceptance is normalised to unit area to allow a comparison of the shape of the distributions.  }
  \label{fig:acceptance}
\end{figure}

At small values of \qsq, there is a large reduction of the signal efficiency
 at values of $\cos\theta_l$ close to $\pm 1$, as seen in Fig.~\ref{fig:acceptance}. 
This results from the requirement for muons to have $p \gsim 3\gevc$ to
reach the  muon system.  Smaller reductions of the signal efficiency
 also arise from the \pt requirement of the hardware
trigger and the impact parameter requirements on the $\mu^\pm$ in the selection.

For the decay \decay{\Bp}{\Kp\mumu}, the \Dzb veto described in
Sect.~\ref{sec:Selection} introduces an additional bias to the angular acceptance: at a
fixed value of \qsq, there is a one-to-one correspondence between
$\cos\theta_l$ and the reconstructed \Dzb mass, and the \Dzb veto
therefore removes a narrow region of $\cos\theta_l$ in each \qsq
bin. The \Dzb veto results in the dip in the acceptance seen in Fig.~\ref{fig:acceptance}.
 The impact of the veto is approximated as a step function in the acceptance model
  and determined using a SM-like sample of simulated events.

\section{Angular analysis} 
\label{sec:Angular} 

\begin{figure}[!tb]
\centering 
\includegraphics[scale=0.38]{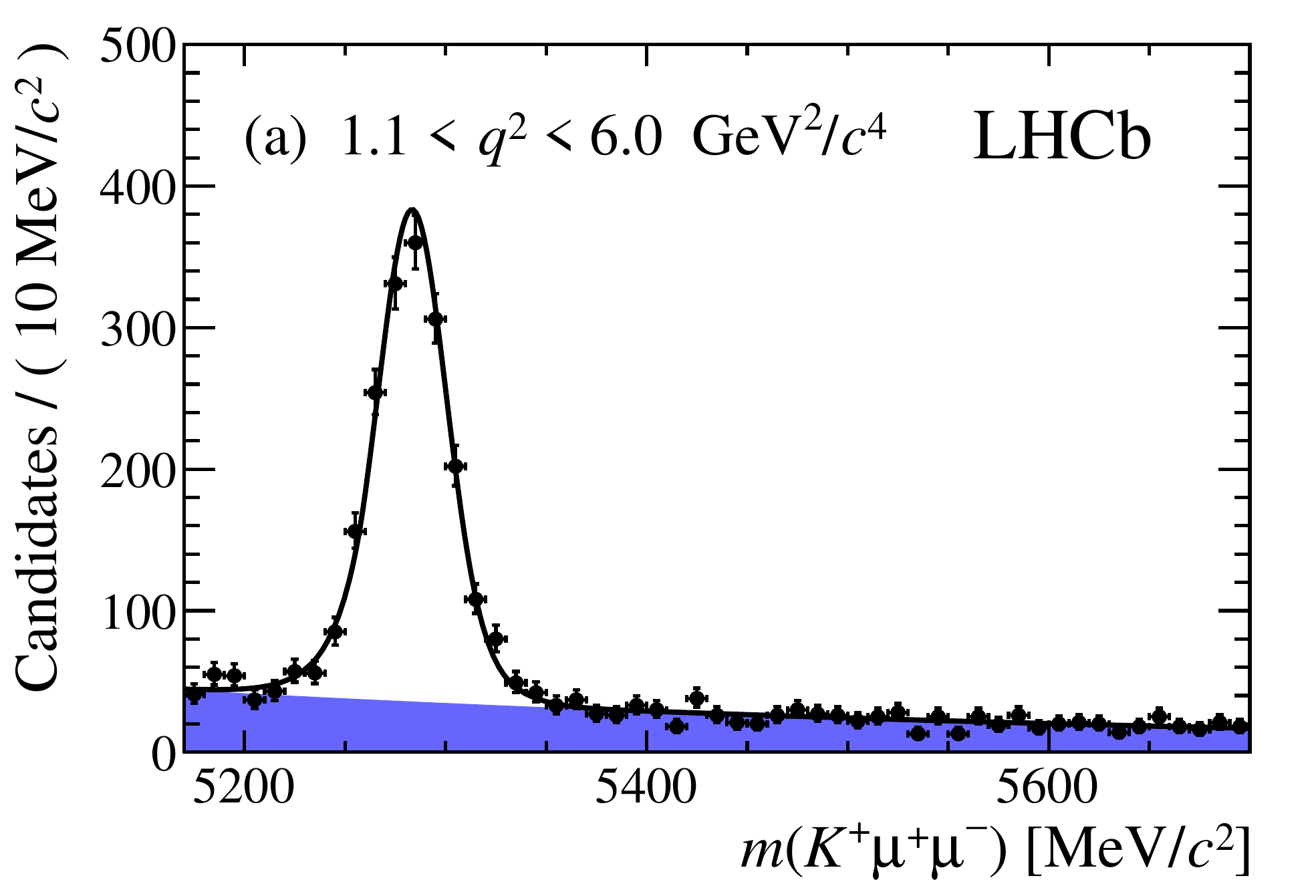} 
\includegraphics[scale=0.38]{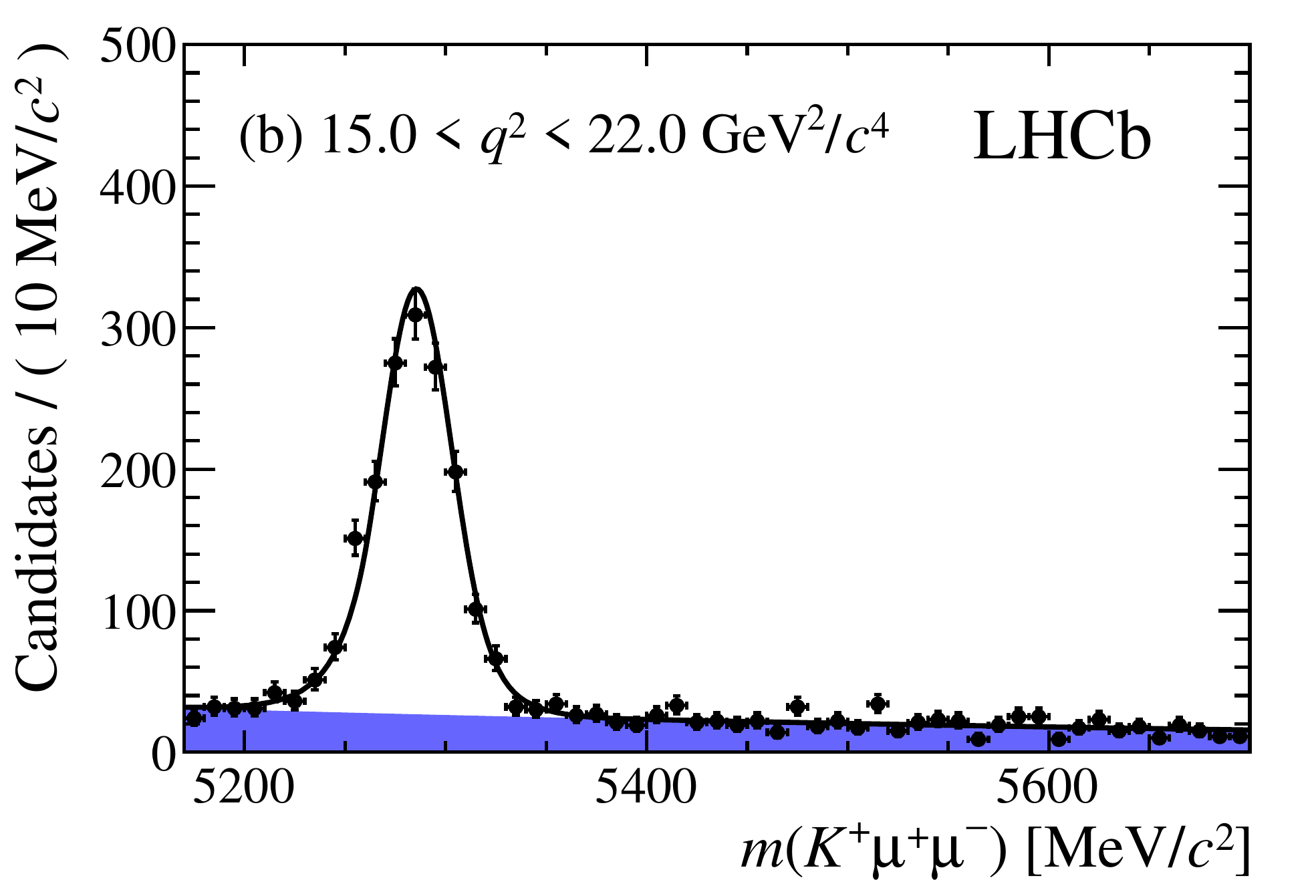} \\ 
\includegraphics[scale=0.38]{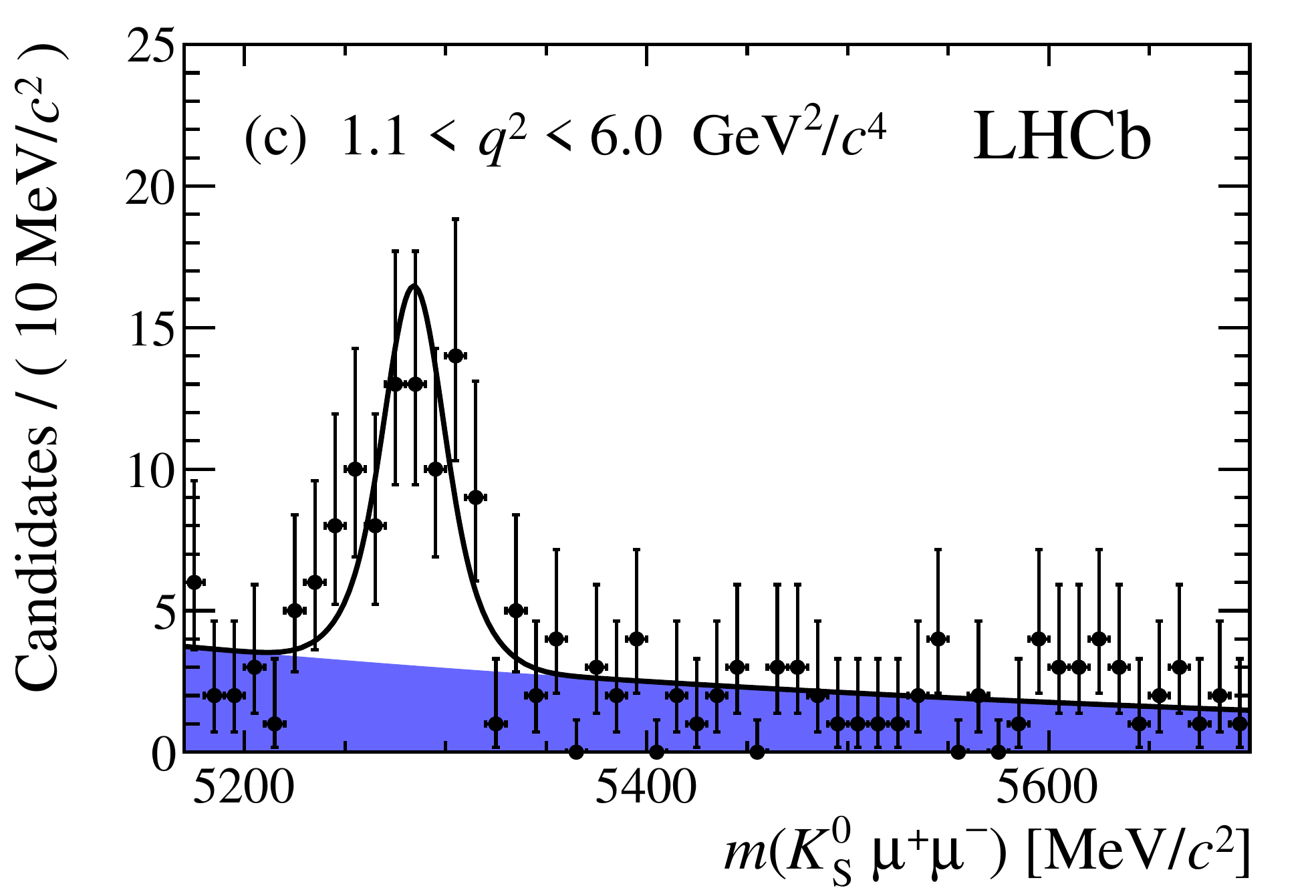} 
\includegraphics[scale=0.38]{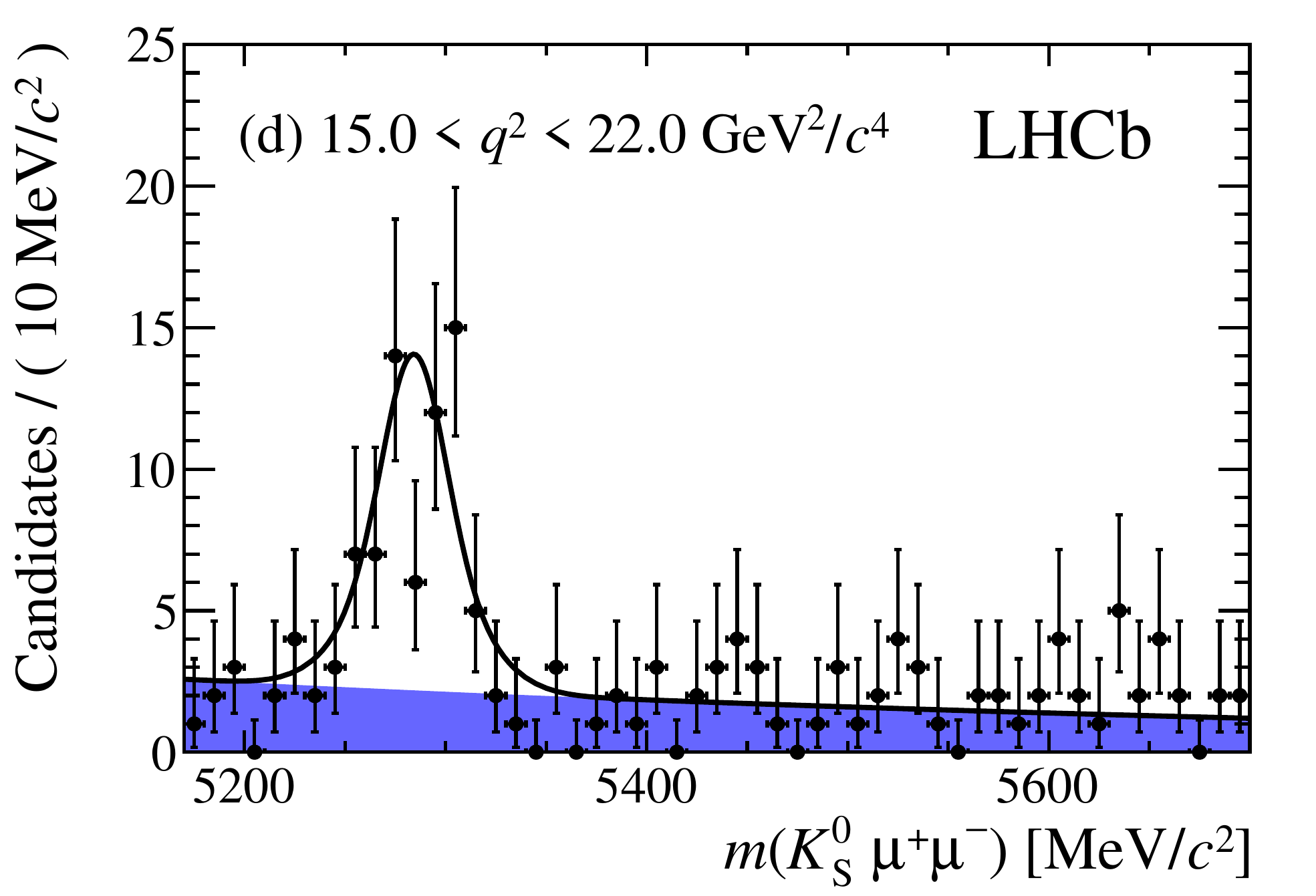} 
\caption{Top, reconstructed mass of \decay{\Bp}{\Kp\mumu} candidates in the ranges (a) $1.1 < \qsq < 6.0\gevgevcccc$ and (b) $15.0 < \qsq < 22.0\gevgevcccc$. Bottom, reconstructed mass of \decay{\Bz}{\KS\mumu} candidates in the ranges (c) $1.1 < \qsq < 6.0\gevgevcccc$ and (d) $15.0 < \qsq < 22.0\gevgevcccc$. The data are overlaid with the result of the fit described in the text. The long and downstream \KS categories are combined for presentation purposes. The shaded region indicates the background contribution in the fit.  \label{fig:Angular:mass}}
\end{figure}
\begin{figure}[!tb]
\centering 
\includegraphics[scale=0.38]{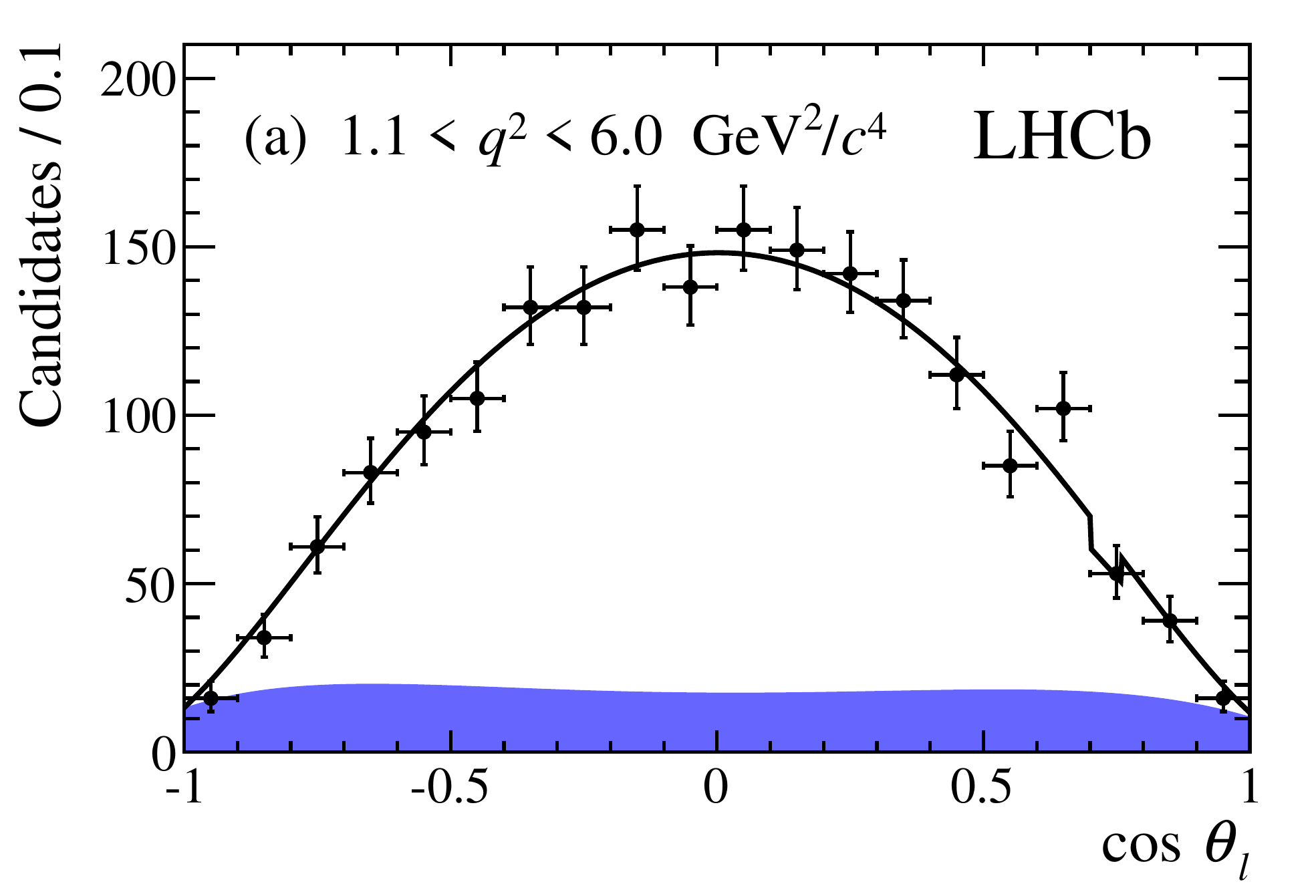} 
\includegraphics[scale=0.38]{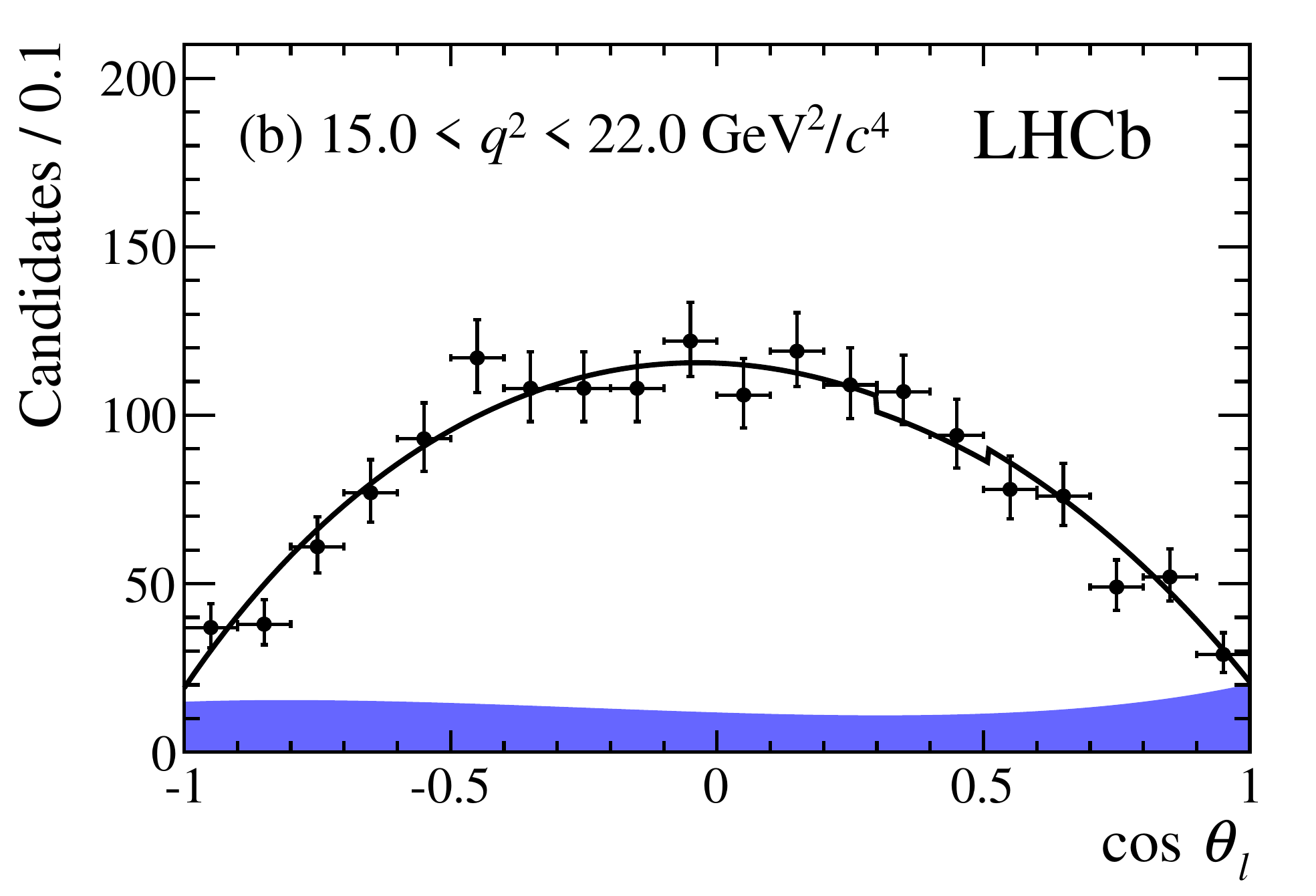} \\ 
\includegraphics[scale=0.38]{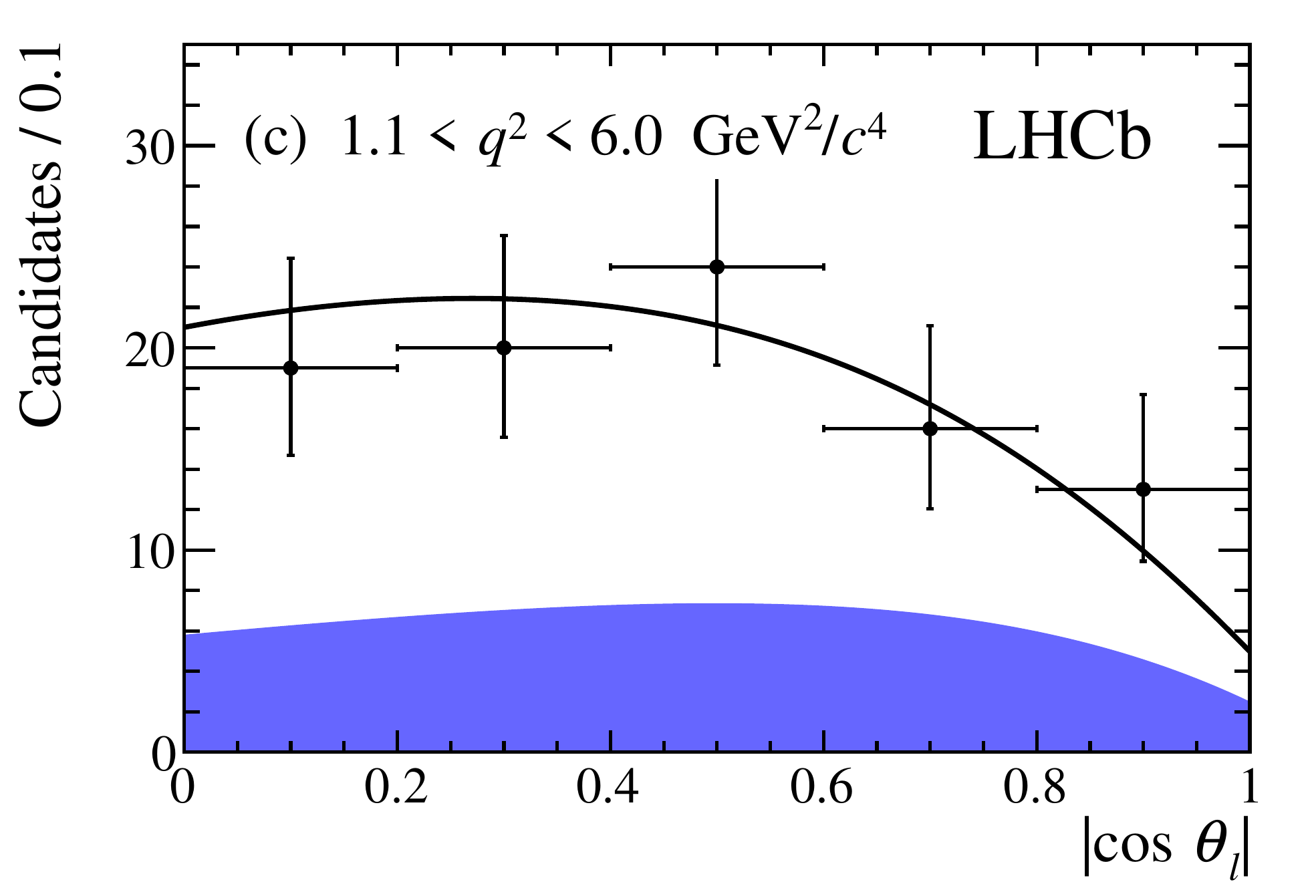} 
\includegraphics[scale=0.38]{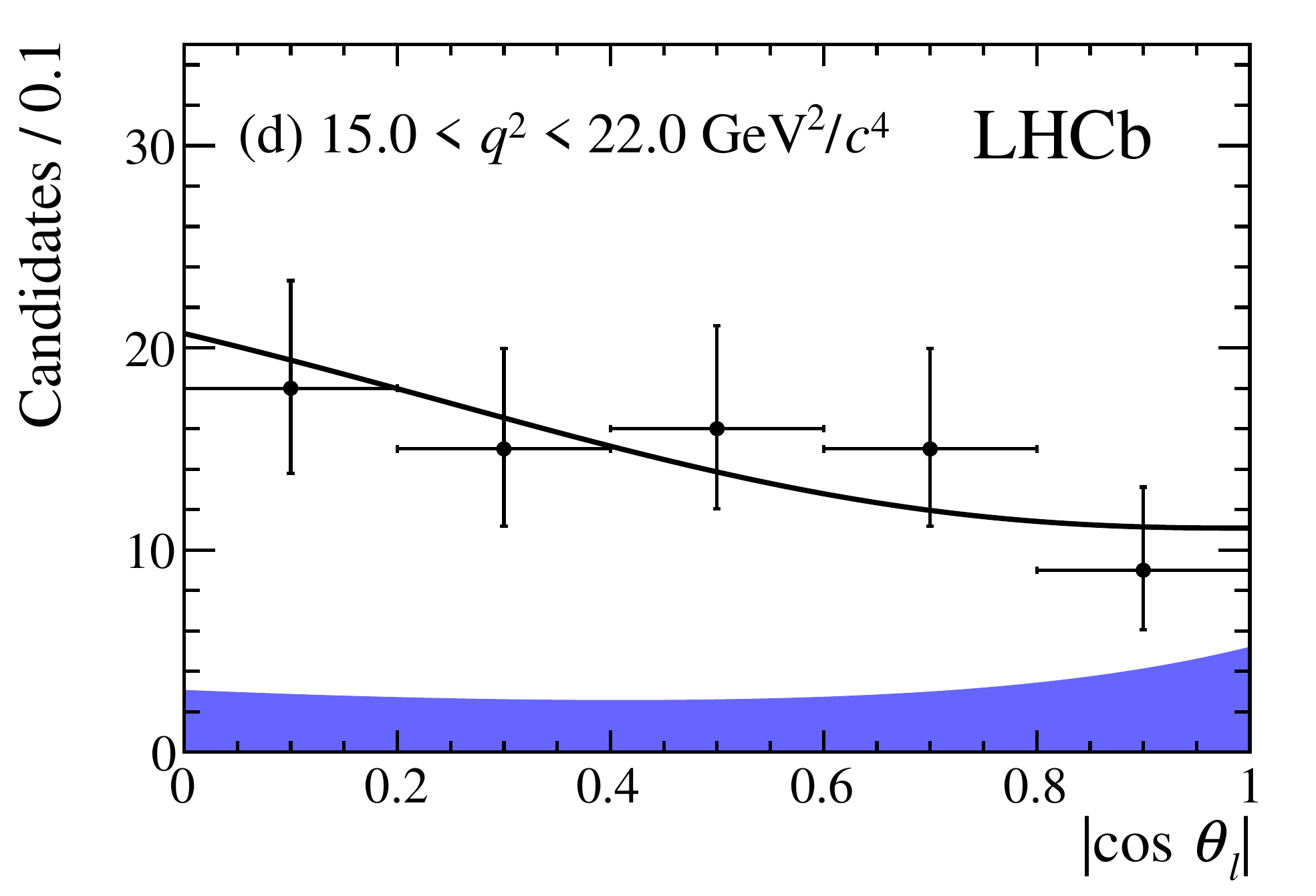} 
\caption{Top, angular distribution of \decay{\Bp}{\Kp\mumu} candidates with (a) $1.1 < \qsq < 6.0\gevgevcccc$ and (b) $15.0 < \qsq < 22.0\gevgevcccc$. Bottom, angular distribution of \decay{\Bz}{\KS\mumu} candidates with (c) $1.1 < \qsq < 6.0\gevgevcccc$ and (d) $15.0 < \qsq < 22.0\gevgevcccc$. Only candidates with a reconstructed mass within $\pm50\mevcc$ of the known \Bp or \Bz mass are shown. The data are overlaid with the result of the fit described in the text. The long and downstream \KS categories are combined for presentation purposes. The shaded region indicates the background contribution in the fit.  \label{fig:Angular:costhetal}}
\end{figure}

The $m(\Kp\mumu)$ and $m(\KS\mumu)$ invariant mass distributions of
candidates that pass the full selection procedure are shown in
Fig.~\ref{fig:Angular:mass}, for two \qsq intervals. The long and downstream categories are
combined for the decay \decay{\Bz}{\KS\mumu}. The angular
 distribution of the candidates is shown in Fig.~\ref{fig:Angular:costhetal}. 

For the \decay{\Bp}{\Kp\mumu} decay, $A_{\rm FB}$ and $F_{\rm H}$ are
determined by performing an unbinned maximum likelihood
fit to the $m(\Kp\mumu)$ and $\cos\theta_l$ distributions of the candidates in bins of \qsq. The signal
angular distribution is described by Eq.~\ref{eq:angular:Bp},
multiplied by the acceptance distribution described in
Sec.~\ref{sec:Acceptance}. The signal mass distribution is
parameterised by the sum of two Gaussian functions with power-law tails, with common most probable values and common tail parameters,
but different widths. The parameters of the these signal functions are
obtained fitting the $m(\Kp\mumu)$ distribution of
\decay{\Bp}{\jpsi\Kp} candidates in data. The peak position and
width parameters are then corrected, using simulated events,
to account for kinematic differences between the decays
\decay{\Bp}{\Kp\mumu} and \decay{\Bp}{\jpsi\Kp}. The $m(\Kp\mumu)$ distribution of the combinatorial background is
parameterised by a falling exponential function. Its angular
distribution is parameterised by a third-order polynomial function multiplied
by the same angular acceptance function used for the signal. 


Decays of \Bz and \Bzb mesons to the $\KS\mumu$ final state cannot 
be separated based on the final-state particles.  The angular distribution of $|\!\cos\theta_l|$ is described by
Eq.~\ref{eq:angular:B0}, which depends only on $F_{\rm
  H}$. Simultaneous unbinned maximum likelihood fits are then performed to
the $|\!\cos\theta_l|$ and $m(\KS\mumu)$ distributions of the two
categories of \KS meson (long and downstream).  The only parameter that is common between
the two simultaneous fits is $F_{\rm H}$. The
$m(\KS\mumu)$ shape parameters of the two categories 
are determined in the same way as that of the decay
\decay{\Bp}{\Kp\mumu}, using \decay{\Bz}{\jpsi\KS} decays. 
Information on the angular shape of the background in the likelihood fit
 is obtained from the upper mass sideband, $5350 < m(\KS\mumu) < 5700\mevcc$. 
 For candidates in the long \KS category, the number of candidates in the
  sideband is so small that the shape is assumed to be uniform. For the downstream category, 
  the shape is parameterised by a second-order polynomial.  The signal and background
angular distributions are then both multiplied by the signal angular
acceptance distribution. The $m(\KS\mumu)$ distribution of the
background candidates is parameterised by a falling exponential function.

The likelihood fits for the \decay{\Bp}{\Kp\mumu} decay and the two categories
 of \KS meson in the \decay{\Bz}{\KS\mumu} decay are performed in two 
 dimensions, treating $m(\Kp\mumu)$ and $\cos\theta_l$ as independent 
 variables.  In total, there are
$4746 \pm 81$ reconstructed signal candidates for the \decay{\Bp}{\Kp\mumu}
decay and $176 \pm 17$ for the \decay{\Bz}{\KS\mumu} decay, summing the yields of the individual \qsq bins.

\section{Results} 
\label{sec:Results} 

\begin{figure}[!tb]
  \centering
  \includegraphics[scale=0.38]{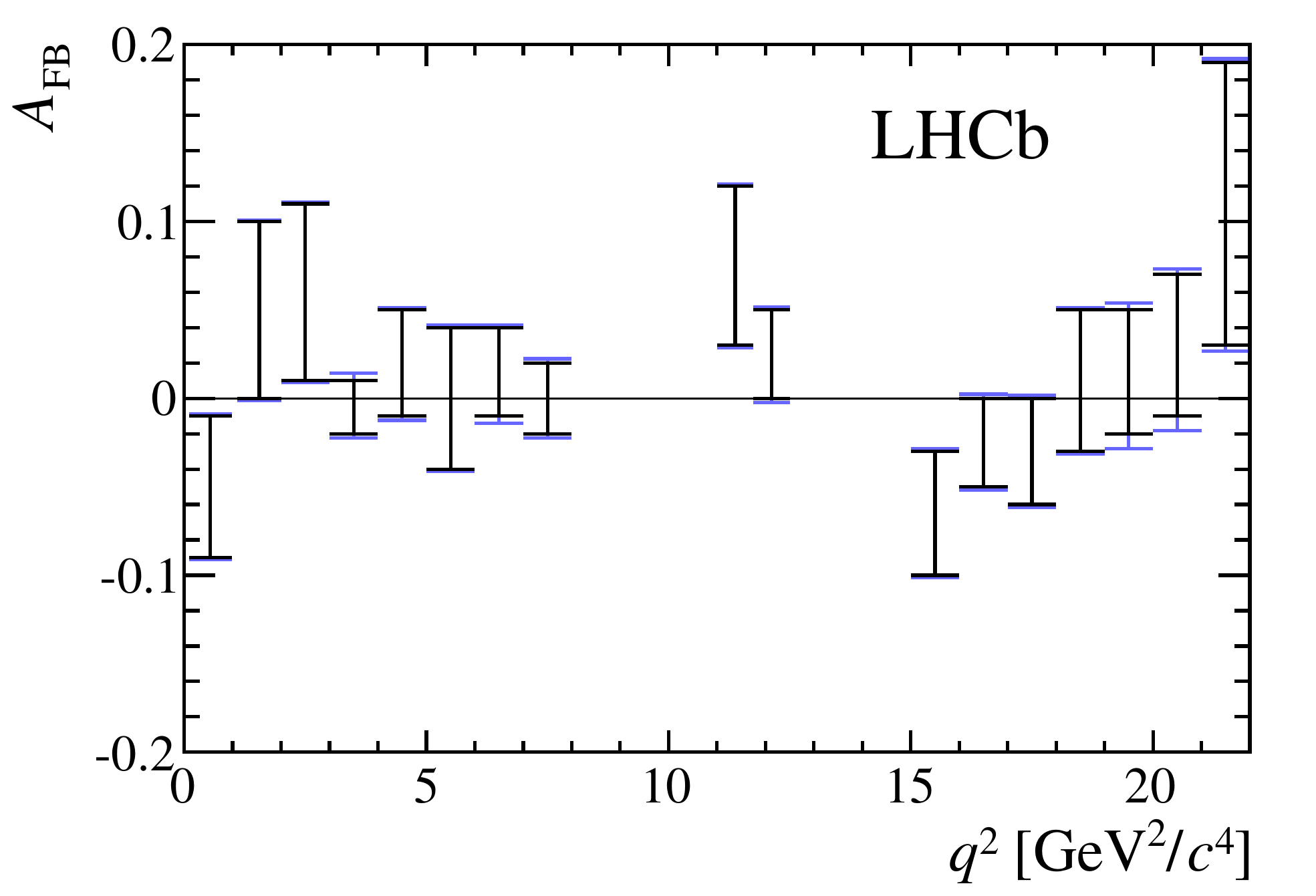} 
  \includegraphics[scale=0.38]{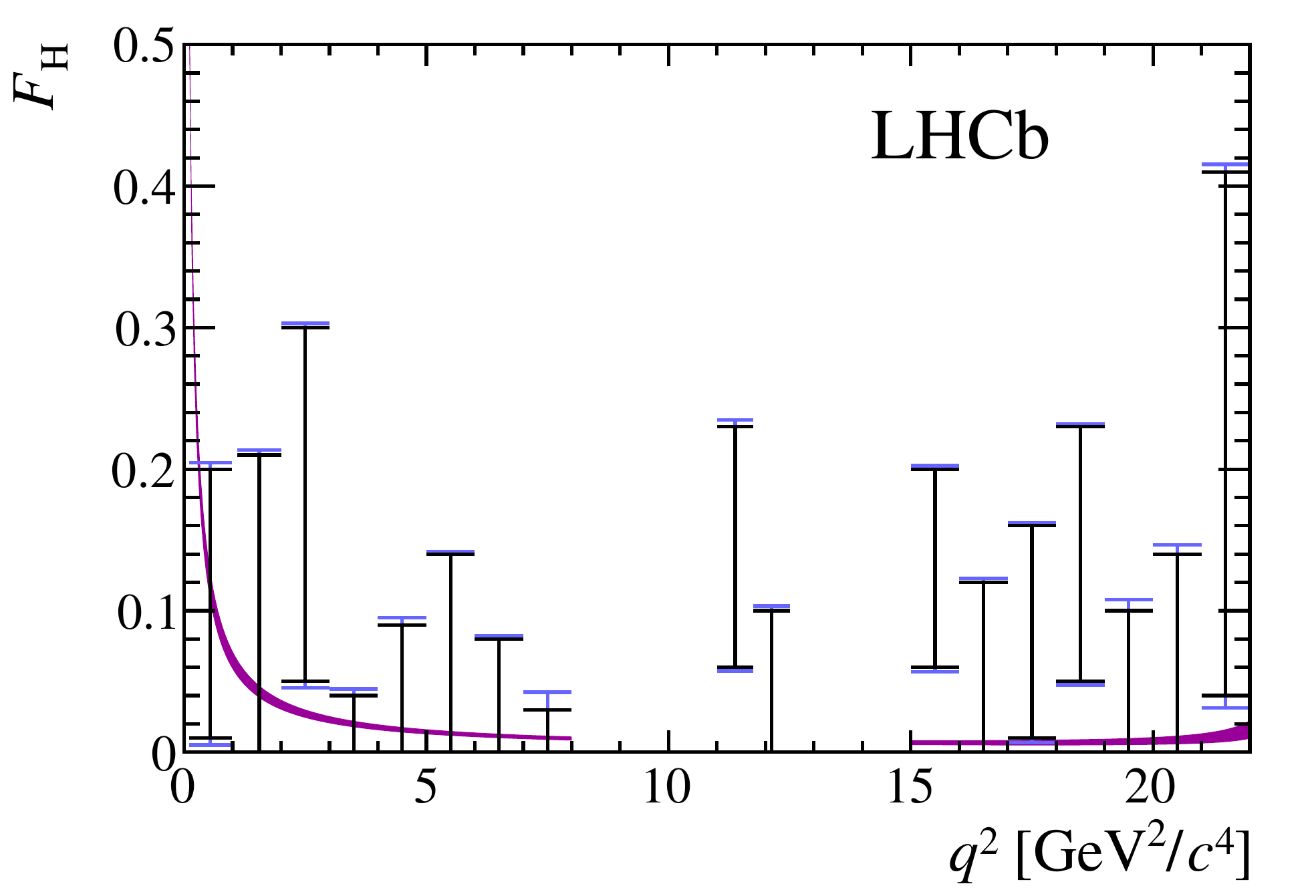}
  \caption{Dimuon forward-backward asymmetry, $A_{\rm FB}$, and the parameter 
    $F_{\rm H}$ for the decay \decay{\Bp}{\Kp\mumu} as a function of the dimuon 
    invariant mass squared, \qsq. The inner horizontal bars indicate the one-dimensional 68\% confidence intervals. The outer vertical bars include contributions
from systematic uncertainties (described in the text). The confidence intervals for $F_{\rm H}$ are overlaid 
    with the SM theory prediction (narrow band).
    Data are not presented for the regions around the \jpsi and \psitwos resonances. 
    \label{fig:Results:Bp}}
\end{figure}

\begin{figure}[!tb]
  \centering
  \includegraphics[scale=0.38]{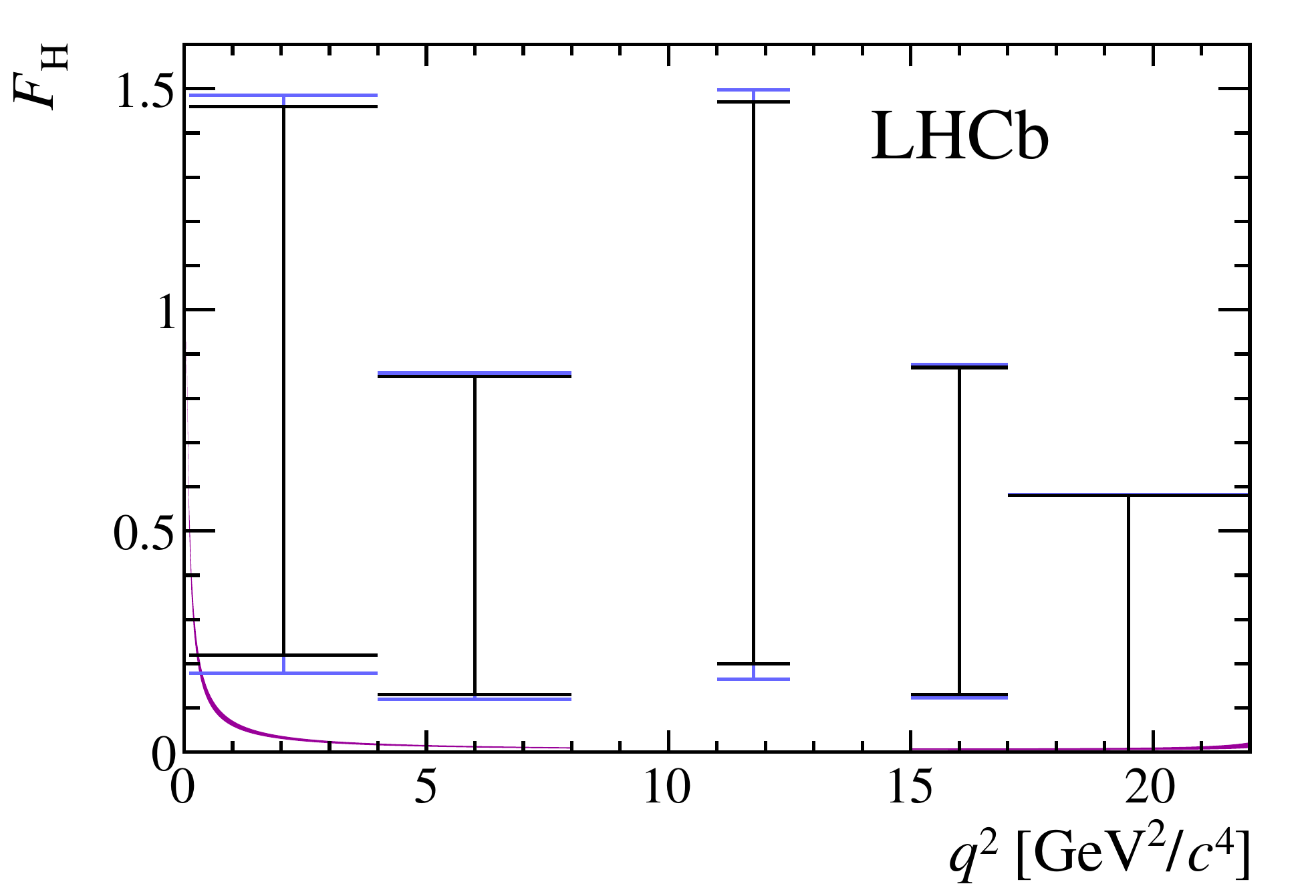}
  \caption{
    Results for the parameter $F_{\rm H}$ for the decay \decay{\Bz}{\KS\mumu} as a function of the dimuon invariant mass squared, \qsq. The inner horizontal bars indicate the one-dimensional 68\% confidence intervals. The outer vertical bars include contributions
from systematic uncertainties (described in the text). The confidence intervals are overlaid with the SM theory prediction (narrow band). Data are not presented for the regions around the \jpsi and \psitwos resonances. 
    \label{fig:Results:B0}
  }
\end{figure}

For the decay \decay{\Bp}{\Kp\mumu}, the results are presented as two-dimensional confidence regions for $A_{\rm FB}$ and $F_{\rm H}$ and as one-dimensional 68\% confidence intervals for $A_{\rm FB}$ and $F_{\rm H}$. The two-dimensional confidence regions demonstrate the correlation between $A_{\rm FB}$ and $F_{\rm H}$ arising from Eq.~\ref{eq:angular:Bp}. The one-dimensional intervals are intended for illustration purposes only. Two-dimensional confidence regions, for the \qsq ranges $1.1 < \qsq < 6.0\gevgevcccc$ and $15.0 < \qsq < 22.0\gevgevcccc$ are shown in Fig.~\ref{fig:Results:FC2D}; the other \qsq bins are provided in the appendix, with the numerical values available from Ref.~\cite{supp}. The one-dimensional confidence intervals for \decay{\Bp}{\Kp\mumu} decays are shown in Fig.~\ref{fig:Results:Bp} and given in Table~\ref{tab:Results:Bp}. The result of the fits to $|\cos\theta_{l}|$ for
the decay \decay{\Bz}{\KS\mumu} are shown in Fig.~\ref{fig:Results:B0}
and given in Table~\ref{tab:Results:B0}. Results are presented in 17 (5) bins of \qsq for the \decay{\Bp}{\Kp\mumu} (\decay{\Bz}{\KS\mumu}) decay. They are also presented in two wide bins of \qsq: one at low hadronic recoil above the open charm threshold and one at large recoil, below the \jpsi meson mass.

The confidence intervals on $F_{\rm H}$ and $A_{\rm FB}$ are estimated using
 the Feldman-Cousins technique~\cite{Feldman:1997qc}. Nuisance parameters are incorporated using the
so-called plug-in method~\cite{woodroofe}. At each value of $F_{\rm H}$ and $A_{\rm FB}$ 
 considered, the maximum likelihood estimate of the nuisance
parameters in data is used when generating the pseudoexperiments. For the 
\decay{\Bp}{\Kp\mumu} decay,  $A_{\rm FB}$ ($F_{\rm H}$) is treated 
 as if it were a nuisance parameter when determining the one-dimensional confidence
 interval on $F_{\rm H}$ ($A_{\rm FB}$). The physical boundaries, described in 
 Sect.~\ref{sec:Introduction}, are accounted for in the generation of pseudoexperiments
  when building the confidence belts. 
 Due to the requirement that $|A_{\rm FB}| \leq F_{\rm H}/2$, statistical fluctuations of
  events in $\cos\theta_l$ have a tendency to drive $F_{\rm H}$ to small positive values in the pseudoexperiments. 

For the \decay{\Bz}{\KS\mumu} decay, fits are also performed to $\cos\theta_{l}$
allowing for a non-zero $A_{\rm FB}$ using Eq.~\ref{eq:angular:Bp}. The value of $A_{\rm FB}$
determined by these fits is consistent with zero, as expected, and the
best fit value of $F_{\rm H}$ compatible with that of the baseline
fit. 

The data for $F_{\rm H}$ in Figs.~\ref{fig:Results:Bp} and \ref{fig:Results:B0}  are superimposed with theoretical predictions from Ref.~\cite{Bobeth:2011nj}. In the low \qsq region, these predictions rely
on the QCD factorisation approaches from 
Ref.~\cite{Bobeth:2007dw}, which lose accuracy when the dimuon mass approaches the 
\jpsi mass. In the high \qsq region, an operator
product expansion in the inverse $b$-quark mass, $1/m_\bquark$, and in
$1/\sqrt{\qsq}$ is used based on Ref.~\cite{Grinstein:2004vb}. This
expansion is only valid above the open charm threshold. A dimensional
estimate of the uncertainty associated with this expansion is
discussed in Ref.~\cite{Egede:2008uy}. Form-factor calculations are taken from
Ref.~\cite{Khodjamirian:2010vf}



\begin{figure}[!tb]
  \centering
  \includegraphics[scale=0.38]{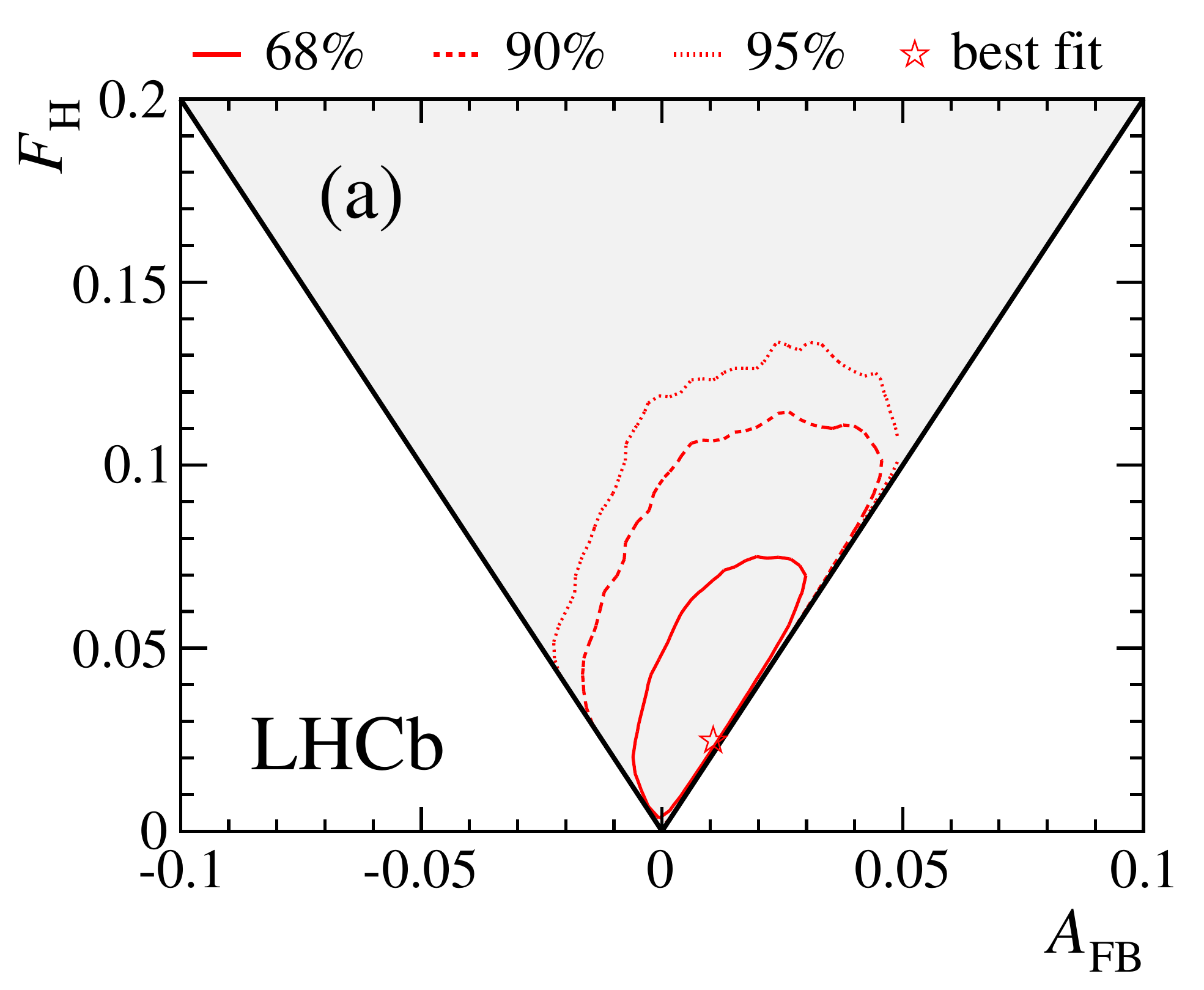} 
  \includegraphics[scale=0.38]{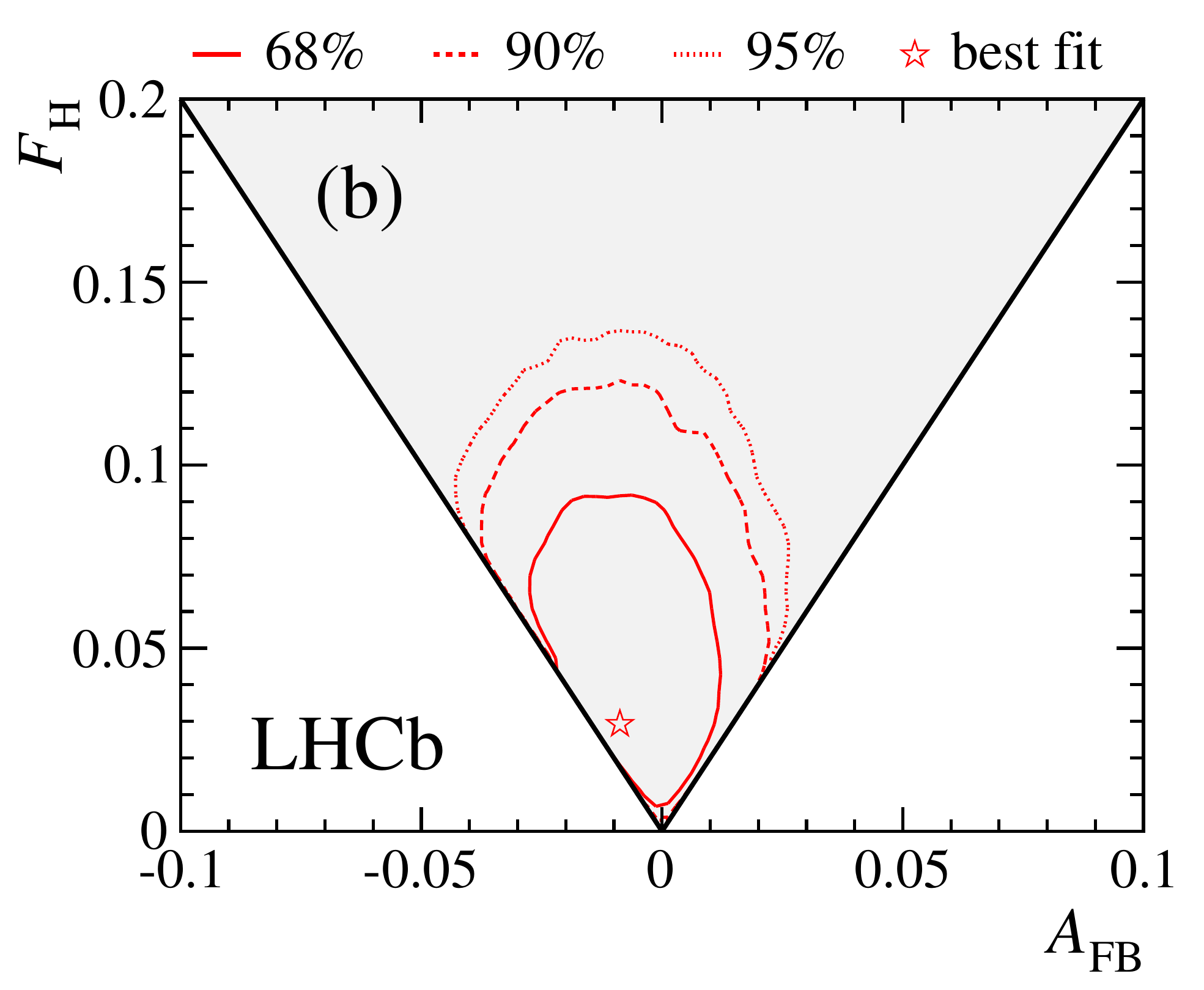}
  \caption{
    Two-dimensional confidence regions for $A_{\rm FB}$ and $F_{\rm H}$ for the decay \decay{\Bp}{\Kp\mumu} in the \qsq ranges (a) $1.1 < \qsq < 6.0\gevgevcccc$ and (b) $15.0 < \qsq < 22.0\gevgevcccc$. The confidence intervals are determined using the Feldman-Cousins technique. The shaded (triangular) region illustrates the range of $A_{\rm FB}$ and $F_{\rm H}$ over which the signal angular distribution remains positive in all regions of phase-space.
    \label{fig:Results:FC2D}}
\end{figure}

\begin{table}[!tb] 
  \caption{
    Forward-backward asymmetry, $A_{\rm FB}$, and $F_{\rm H}$ for the decay 
    \decay{\Bp}{\Kp\mumu} in the \qsq bins used in this analysis.  
    These parameters are also given in a wide bin at large ($1.1 < \qsq  <6.0\gevgevcccc$) and low ($15.0 < \qsq < 22.0\gevgevcccc$) hadronic recoil.  
    The column labelled \emph{stat} is the 68\% statistical confidence interval on $F_{\rm H}$ ($A_{\rm FB}$) when treating $A_{\rm FB}$ ($F_{\rm H}$) as a nuisance parameter. The column labelled \emph{syst} is the systematic uncertainty. 
    \label{tab:Results:Bp}
 }
\centering
\setlength{\extrarowheight}{2pt}
\begin{tabular}{ccccc}
$\qsq (\gev^2/c^4)$ & $F_{\rm H}$ (stat) & $F_{\rm H}$ (syst) & $A_{\rm FB}$ (stat) & $A_{\rm FB}$ (syst)  \\
\hline
$0.10 - 0.98$ & $[+0.01,+0.20]$ & $\pm0.03$ & $[-0.09,-0.01]$ & $\pm0.01$ \\
$1.10 - 2.00$ & $[+0.00,+0.21]$ & $\pm0.03$ & $[+0.00,+0.10]$ & $\pm0.01$ \\
$2.00 - 3.00$ & $[+0.05,+0.30]$ & $\pm0.03$ & $[+0.01,+0.11]$ & $\pm0.01$ \\
$3.00 - 4.00$ & $[\phantom{+}0.00,+0.04]$ & $\pm0.02$ & $[-0.02,+0.01]$ & $\pm0.01$ \\
$4.00 - 5.00$ & $[\phantom{+}0.00,+0.09]$ & $\pm0.03$ & $[-0.01,+0.05]$ & $\pm0.01$ \\
$5.00 - 6.00$ & $[\phantom{+}0.00,+0.14]$ & $\pm0.02$ & $[-0.04,+0.04]$ & $\pm0.01$ \\
$6.00 - 7.00$ & $[\phantom{+}0.00,+0.08]$ & $\pm0.02$ & $[-0.01,+0.04]$ & $\pm0.01$ \\
$7.00 - 8.00$ & $[\phantom{+}0.00,+0.03]$ & $\pm0.03$ & $[-0.02,+0.02]$ & $\pm0.01$ \\
$11.00 - 11.75$ & $[+0.06,+0.23]$ & $\pm0.03$ & $[+0.03,+0.12]$ & $\pm0.01$ \\
$11.75 - 12.50$ & $[+0.00,+0.10]$ & $\pm0.02$ & $[+0.00,+0.05]$ & $\pm0.01$ \\
$15.00 - 16.00$ & $[+0.06,+0.20]$ & $\pm0.02$ & $[-0.10,-0.03]$ & $\pm0.01$ \\
$16.00 - 17.00$ & $[+0.00,+0.12]$ & $\pm0.02$ & $[-0.05,+0.00]$ & $\pm0.01$ \\
$17.00 - 18.00$ & $[+0.01,+0.16]$ & $\pm0.02$ & $[-0.06,+0.00]$ & $\pm0.01$ \\
$18.00 - 19.00$ & $[+0.05,+0.23]$ & $\pm0.02$  &  $[-0.03,+0.05]$ & $\pm0.01$ \\
$19.00 - 20.00$ & $[\phantom{+}0.00,+0.10]$ & $\pm0.04$ & $[-0.02,+0.05]$ & $\pm0.02$ \\
$20.00 - 21.00$ & $[\phantom{+}0.00,+0.14]$ & $\pm0.04$ & $[-0.01,+0.07]$ & $\pm0.02$ \\
$21.00 - 22.00$ & $[+0.04,+0.41]$ & $\pm0.05$ & $[+0.03,+0.19]$ & $\pm0.02$ \\
\hline
$1.10 - 6.00$ & $[\phantom{+}0.00,+0.06]$ & $\pm0.02$ & $[-0.01,+0.02]$ & $\pm0.01$ \\
$15.00 - 22.00$ & $[\phantom{+}0.00,+0.07]$ & $\pm0.02$ & $[-0.03,+0.00]$ & $\pm0.01$ \\
\end{tabular} 
\end{table}

\begin{table}[!tb] 
  \caption{
    The 68\% confidence interval on the parameter $F_{\rm H}$ for the decay \decay{\Bz}{\KS\mumu} in \qsq 
    bins.  In addition to the narrow binning used in the analysis, results are also given in wide bins at large ($1.1 < \qsq  <6.0\gevgevcccc$) and low ($15.0 < \qsq < 22.0\gevgevcccc$) hadronic recoil. The column labelled \emph{stat} is
     the 68\% statistical confidence interval. The column labelled \emph{syst} is the systematic uncertainty.
\label{tab:Results:B0}
}
\centering
\setlength{\extrarowheight}{2pt}
\begin{tabular}{ccc}
$\qsq (\gev^2/c^4)$ & $F_{\rm H}$ (stat) & $F_{\rm H}$ (syst) \\
\hline
$0.1 - 4.0$ & $[+0.22,+1.46]$ & $\pm0.28$ \\ 
$4.0 - 8.0$ & $[+0.13,+0.85]$ & $\pm0.08$ \\
$11.0 - 12.5$ & $[+0.20,+1.47]$ & $\pm0.20$ \\ 
$15.0 - 17.0$ & $[+0.12,+0.77]$ & $\pm0.07$ \\ 
$17.0 - 22.0$ & $[\phantom{+}0.00,+0.58]$ & $\pm0.04$ \\ 
\hline
$1.1 - 6.0$ & $[+0.32,+1.24]$ & $\pm0.09$ \\ 
$15.0 - 22.0$ & $[+0.09,+0.59]$ & $\pm0.03$ \\
\end{tabular} 
\end{table}

Two classes of systematic uncertainty are considered for $A_{\rm FB}$
and $F_{\rm H}$: detector-related uncertainties that might affect the
angular acceptance, and uncertainties related to the angular distribution of the background. 


The samples of simulated events used to determine the detector
acceptance are corrected to match the performance observed in data by degrading the impact
parameter resolution on the kaon and muons by 20\%, re-weighting
candidates to reproduce the kinematic distribution of \Bp candidates
in the data and re-weighting candidates to account for differences
in tracking and particle-identification performance. Varying these
corrections within their known uncertainties has a negligible impact
on $A_{\rm FB}$ and $F_{\rm H}$ $(\lsim 0.01)$.

The acceptance as a function of $\cos\theta_l$ is determined from
simulated events in each bin of \qsq. This assumes that the
distribution of events in \qsq, within the \qsq bin, is the same in
simulation and in data. To assess the systematic uncertainty arising
from this assumption, the acceptance as a function of $\cos\theta_l$
is determined separately for simulated events in the lower and
upper half of the \qsq bin, and the average acceptance correction for
the bin is re-computed varying the relative contributions from the
lower and upper half by 20\%. This level of variation covers any
observed difference between the differential decay rate as a function
of \qsq in data and in simulation and introduces an uncertainty at the
level of 0.01 on $A_{\rm FB}$ and $F_{\rm H}$.

In order to investigate the background modelling, the multivariate
selection requirements are relaxed. With the increased level of
background in the upper mass sideband, an alternative background model of a fourth-order polynomial is derived. Pseudoexperiments are then generated that explore the differences between the $A_{\rm FB}$ or $F_{\rm H}$ values obtained with the default and
the alternative background model. A systematic uncertainty is assigned based on
the sum in quadrature of the root-mean-square of these differences and the
mean bias. The method introduces an uncertainty at the level of
$0.01$ on $A_{\rm FB}$ and $0.02 - 0.05$ on $F_{\rm H}$ for the
\decay{\Bp}{\Kp\mumu} decay and $0.04 - 0.20$ for the
\decay{\Bz}{\KS\mumu} decay.

The dependence of the one-dimensional $A_{\rm FB}$ ($F_{\rm H}$) confidence interval on the assumed true value of the $F_{\rm H}$ ($A_{\rm FB}$) nuisance parameter is negligible ($\lsim 0.01$).


The fitting procedure for \decay{\Bp}{\Kp\mumu}
(\decay{\Bz}{\KS\mumu}) decays is also tested using samples of
\decay{\Bp}{\jpsi\Kp} (\decay{\Bz}{\jpsi\KS}) decays where $A_{\rm FB} = F_{\rm H} = 0$, due to the vector nature of the \jpsi meson.  These samples are more than one hundred times
larger than the signal samples. Tests are also performed splitting these samples
into sub-samples of comparable size to the data sets in the individual \qsq bins. No indication of any bias is seen in
the fitting procedure in either set of tests.

\section{Conclusion} 
\label{sec:Summary} 

In summary, the angular distributions of charged and neutral \decay{\B}{\kaon \mumu} decays are studied using a data set, corresponding to an integrated luminosity of 3\invfb, collected by the LHCb experiment. The angular distribution of the decays is parameterised in terms of the forward-backward asymmetry of the decay, $A_{\rm FB}$, and a parameter $F_{\rm H}$, which is a measure of the contribution from (pseudo)scalar and tensor amplitudes to the decay width.  

The measurements of $A_{\rm FB}$ and $F_{\rm H}$ presented for the decays \decay{\Bp}{\Kp\mumu} and \decay{\Bz}{\KS\mumu} are the most precise to date. They are consistent with SM predictions ($A_{\rm FB} \approx 0$ and $F_{\rm H} \approx 0$) in every bin of \qsq.  The results are also compatible between the decays \decay{\Bp}{\Kp\mumu} and \decay{\Bz}{\KS\mumu}. The largest difference with respect to the SM prediction is seen in the range $11.00 < \qsq < 11.75\gev^{2}/c^{4}$ for the decay \decay{\Bp}{\Kp\mumu}. Even in this bin, the SM point is included at 95\% confidence level when taking into account the systematic uncertainties on the angular observables.

The results place constraints on (pseudo)scalar and tensor amplitudes, which are vanishingly small in the SM but can be enhanced in many extensions of the SM. Pseudoscalar and scalar amplitudes were already highly constrained by measurements of the branching fraction of the decay \decay{\Bs}{\mumu}~\cite{LHCb-PAPER-2013-046,Chatrchyan:2013bka}. The results presented here, however, also rule out the possibility of large accidental cancellations between the left- and right-handed couplings of the (pseudo)scalar amplitudes to the \decay{\Bs}{\mumu} branching fraction. Tensor amplitudes were previously poorly constrained.

\section*{Acknowledgements}

\noindent We express our gratitude to our colleagues in the CERN
accelerator departments for the excellent performance of the LHC. We
thank the technical and administrative staff at the LHCb
institutes. We acknowledge support from CERN and from the national
agencies: CAPES, CNPq, FAPERJ and FINEP (Brazil); NSFC (China);
CNRS/IN2P3 and Region Auvergne (France); BMBF, DFG, HGF and MPG
(Germany); SFI (Ireland); INFN (Italy); FOM and NWO (The Netherlands);
SCSR (Poland); MEN/IFA (Romania); MinES, Rosatom, RFBR and NRC
``Kurchatov Institute'' (Russia); MinECo, XuntaGal and GENCAT (Spain);
SNSF and SER (Switzerland); NASU (Ukraine); STFC and the Royal Society (United
Kingdom); NSF (USA). We also acknowledge the support received from EPLANET, 
Marie Curie Actions and the ERC under FP7. 
The Tier1 computing centres are supported by IN2P3 (France), KIT and BMBF (Germany),
INFN (Italy), NWO and SURF (The Netherlands), PIC (Spain), GridPP (United Kingdom).
We are indebted to the communities behind the multiple open source software packages on which we depend.
We are also thankful for the computing resources and the access to software R\&D tools provided by Yandex LLC (Russia).


\addcontentsline{toc}{section}{References}
\setboolean{inbibliography}{true}
\ifx\mcitethebibliography\mciteundefinedmacro
\PackageError{LHCb.bst}{mciteplus.sty has not been loaded}
{This bibstyle requires the use of the mciteplus package.}\fi
\providecommand{\href}[2]{#2}



\clearpage

{\noindent\bf\Large Appendix}

\appendix

\label{sec:appendix:intervals} 

\begin{figure}[!h] 
\centering
\subfigure[$0.10 < \qsq < 0.98\gev^{2}/c^{4}$]{\includegraphics[width=0.48\linewidth]{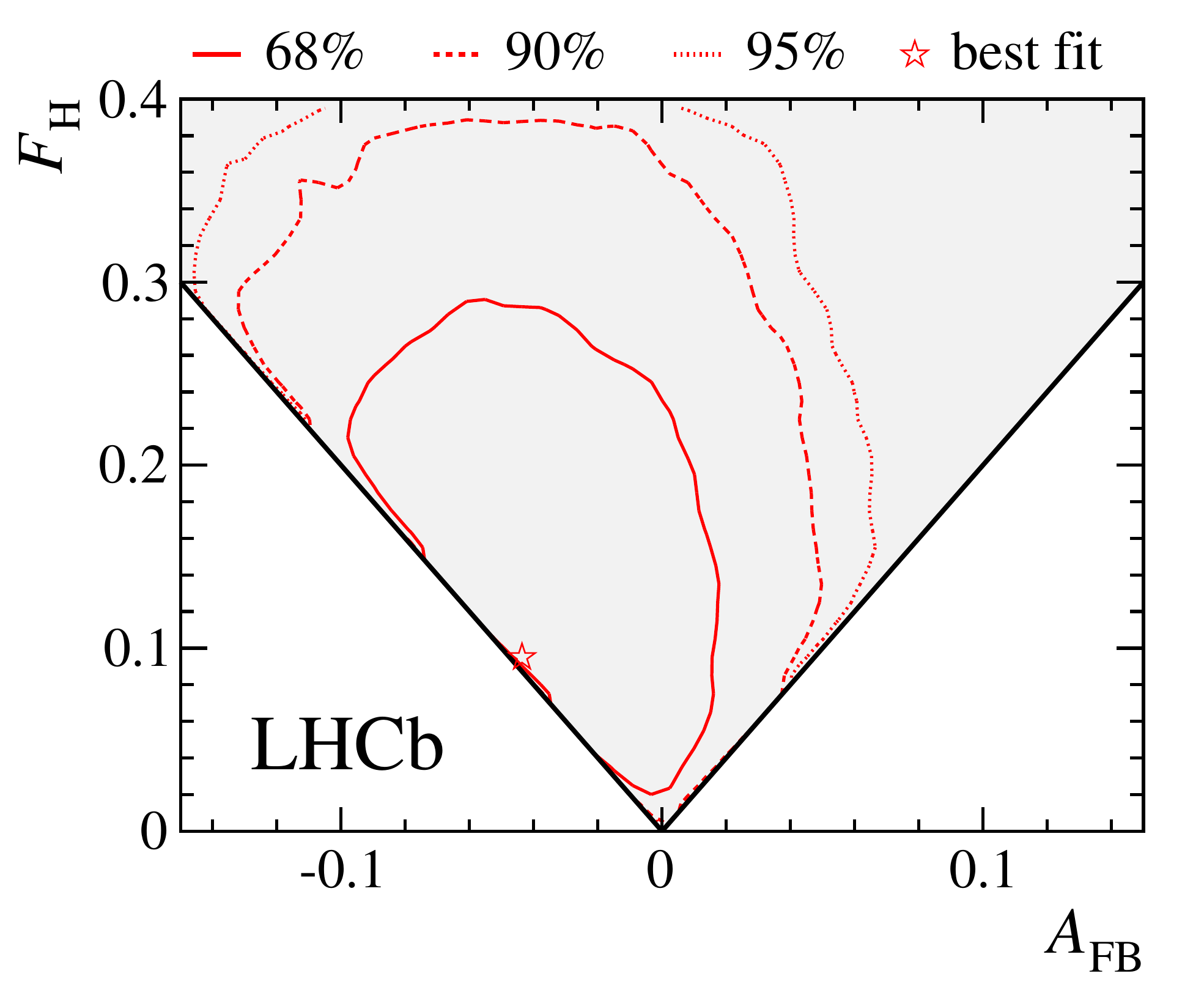}}
\subfigure[$1.10 < \qsq < 2.00\gev^{2}/c^{4}$]{\includegraphics[width=0.48\linewidth]{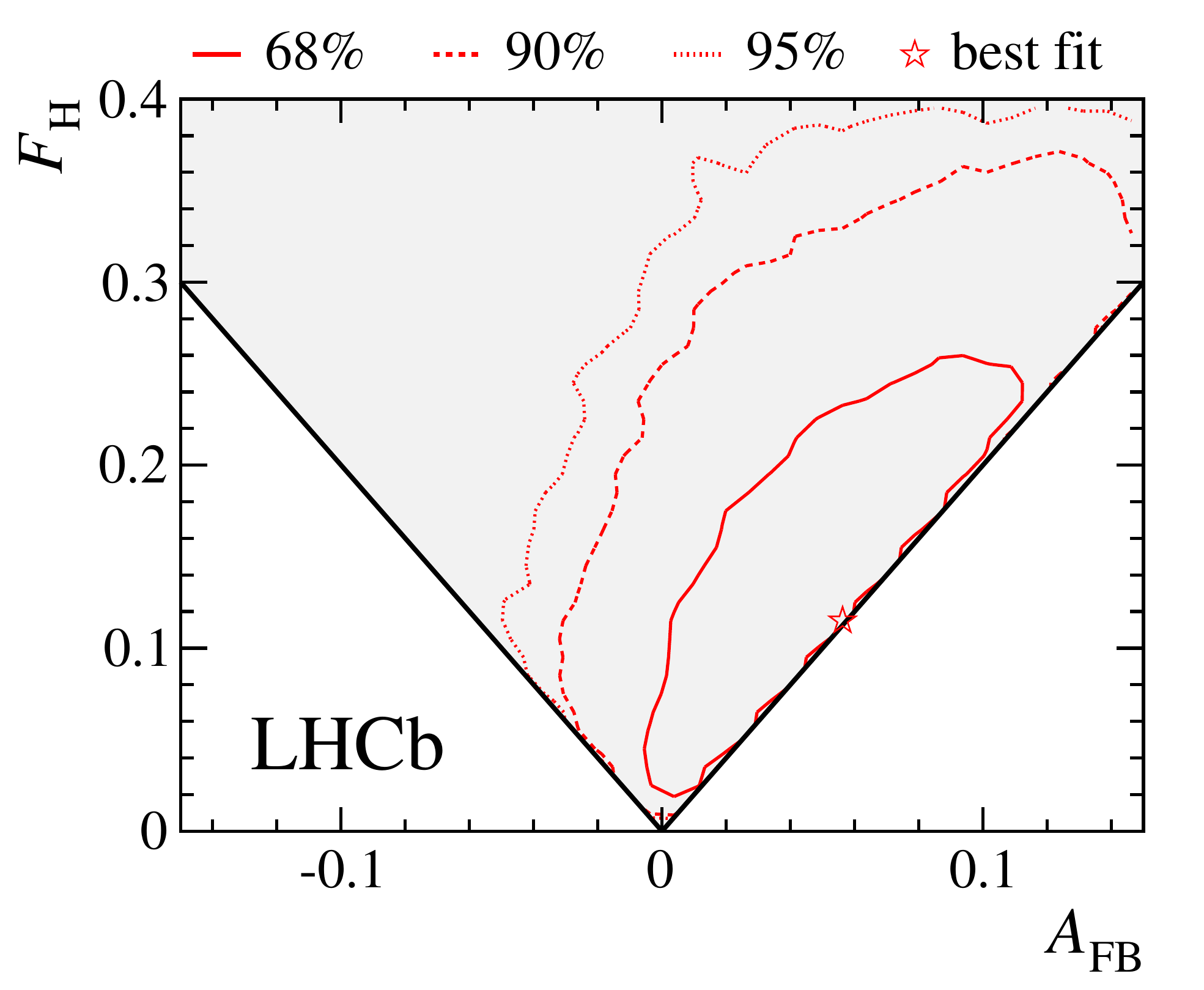}} \\
\subfigure[$2.00 < \qsq < 3.00\gev^{2}/c^{4}$]{\includegraphics[width=0.48\linewidth]{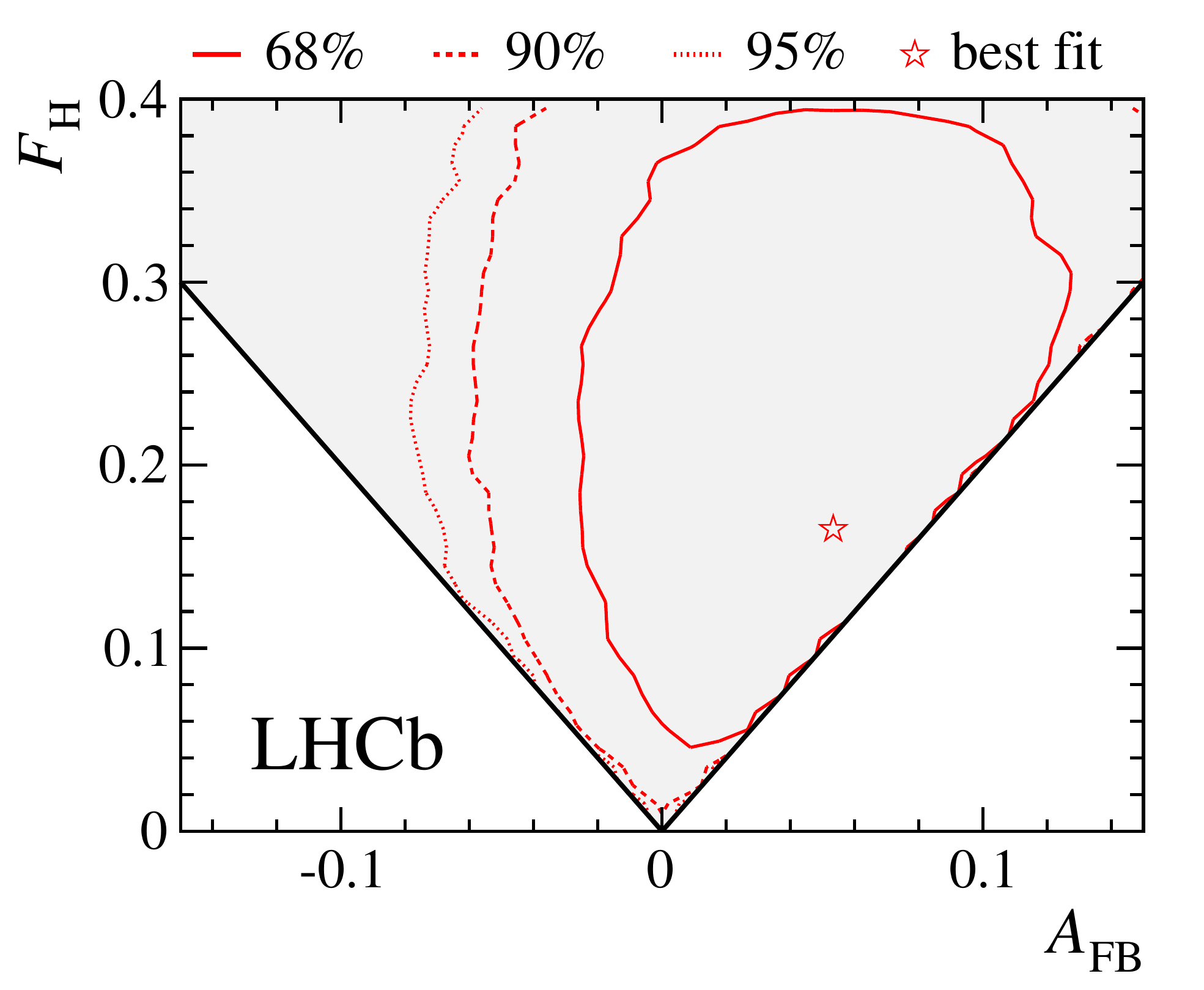}}
\subfigure[$3.00 < \qsq < 4.00\gev^{2}/c^{4}$]{\includegraphics[width=0.48\linewidth]{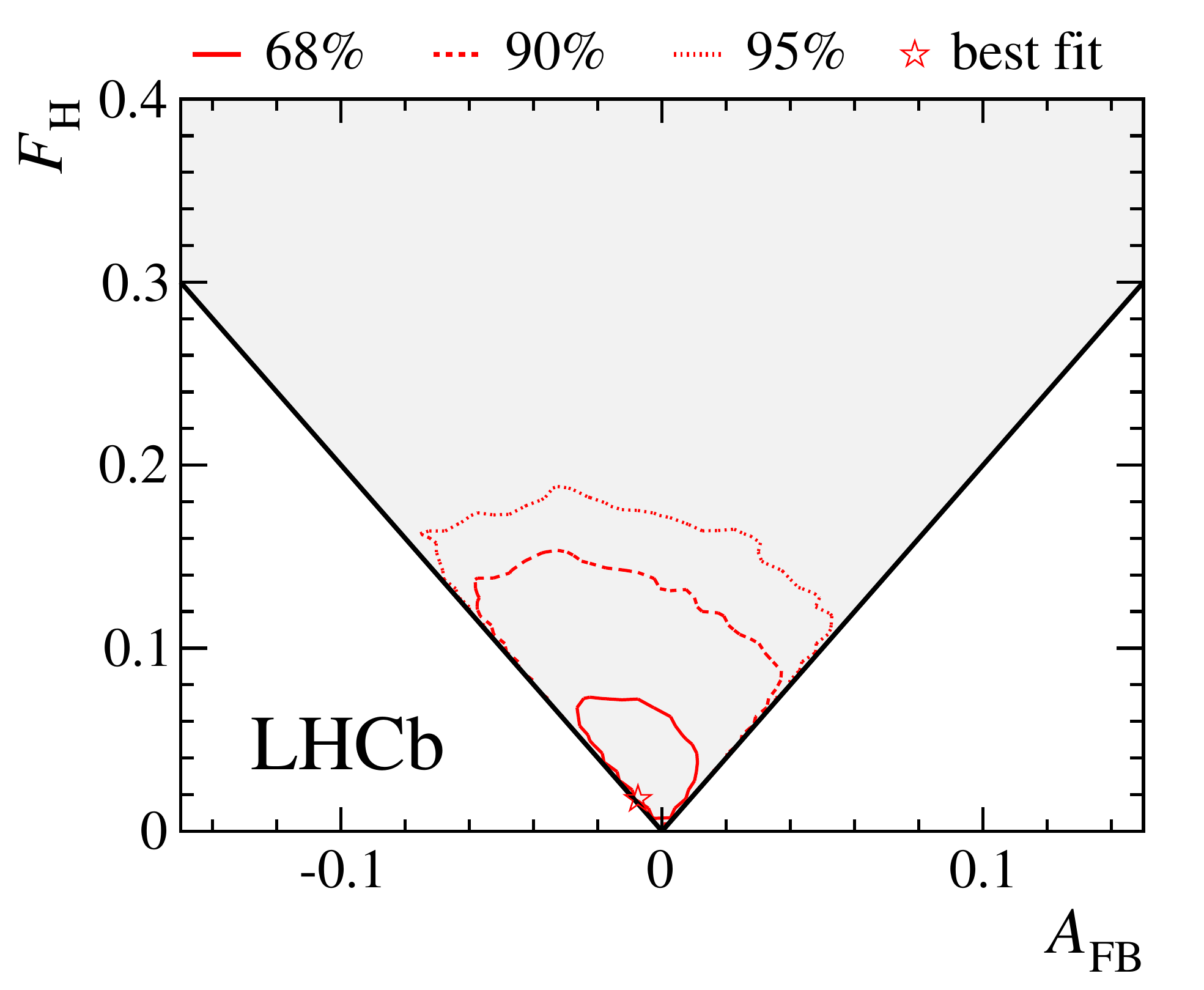}} \\
\caption{Two-dimensional confidence regions for $A_{\rm FB}$ and $F_{\rm H}$ for the decay \decay{\Bp}{\Kp\mumu} in the \qsq ranges (a) $0.10 < \qsq < 0.98\gevgevcccc$, (b) $1.10 < \qsq < 2.00\gevgevcccc$, (c) $2.00 < \qsq < 3.00\gevgevcccc$ and (d) $3.00 < \qsq < 4.00\gevgevcccc$. The confidence intervals are determined using the Feldman-Cousins technique and are purely statistical. The shaded (triangular) region illustrates the range of $A_{\rm FB}$ and $F_{\rm H}$ over which the signal angular distribution remains positive in all regions of phase-space.\label{fig:appendix:2D:A}}
\end{figure}

\begin{figure}[!h] 
\centering
\subfigure[$4.00 < \qsq < 5.00\gev^{2}/c^{4}$]{\includegraphics[width=0.48\linewidth]{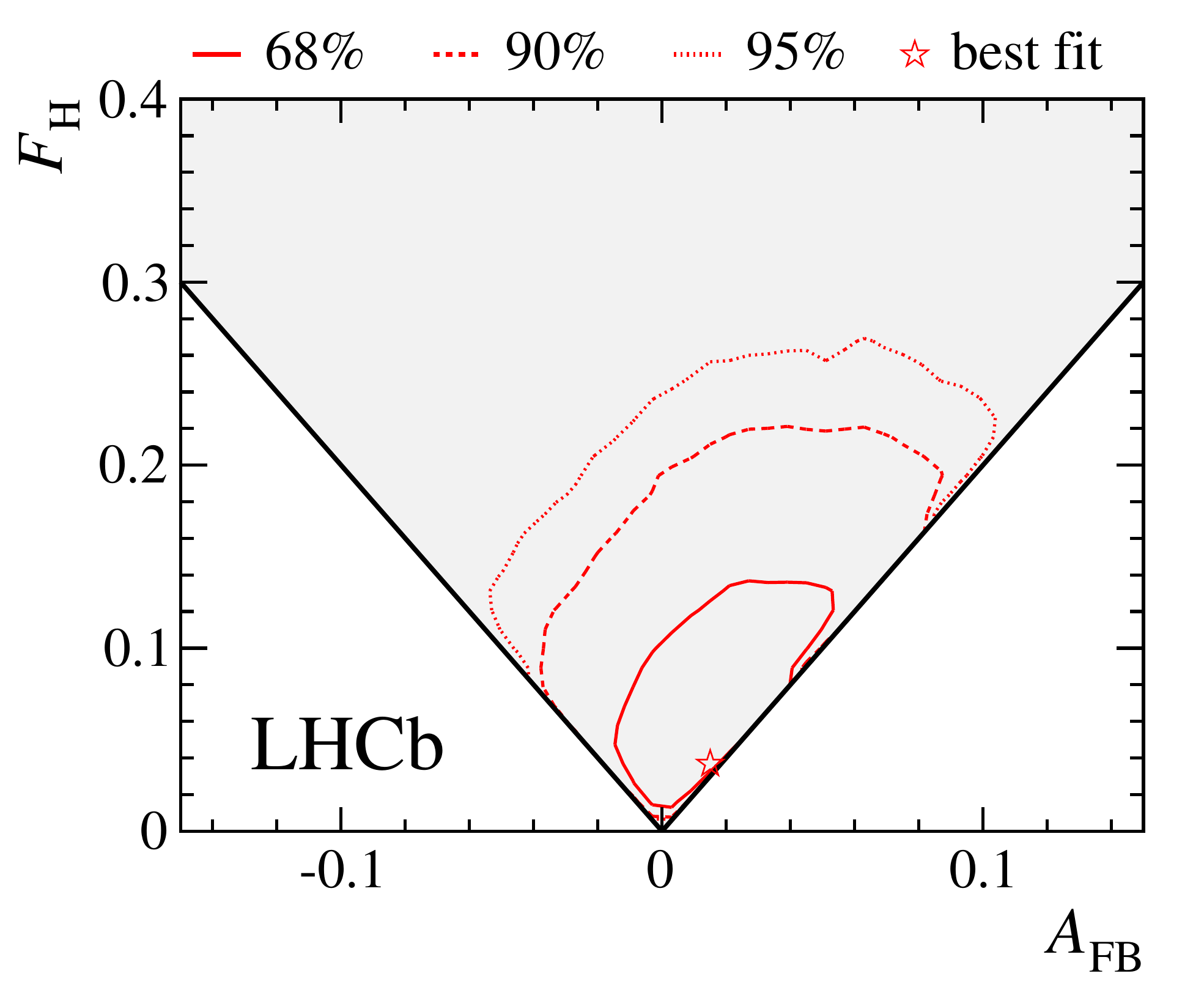}}
\subfigure[$5.00 < \qsq < 6.00\gev^{2}/c^{4}$]{\includegraphics[width=0.48\linewidth]{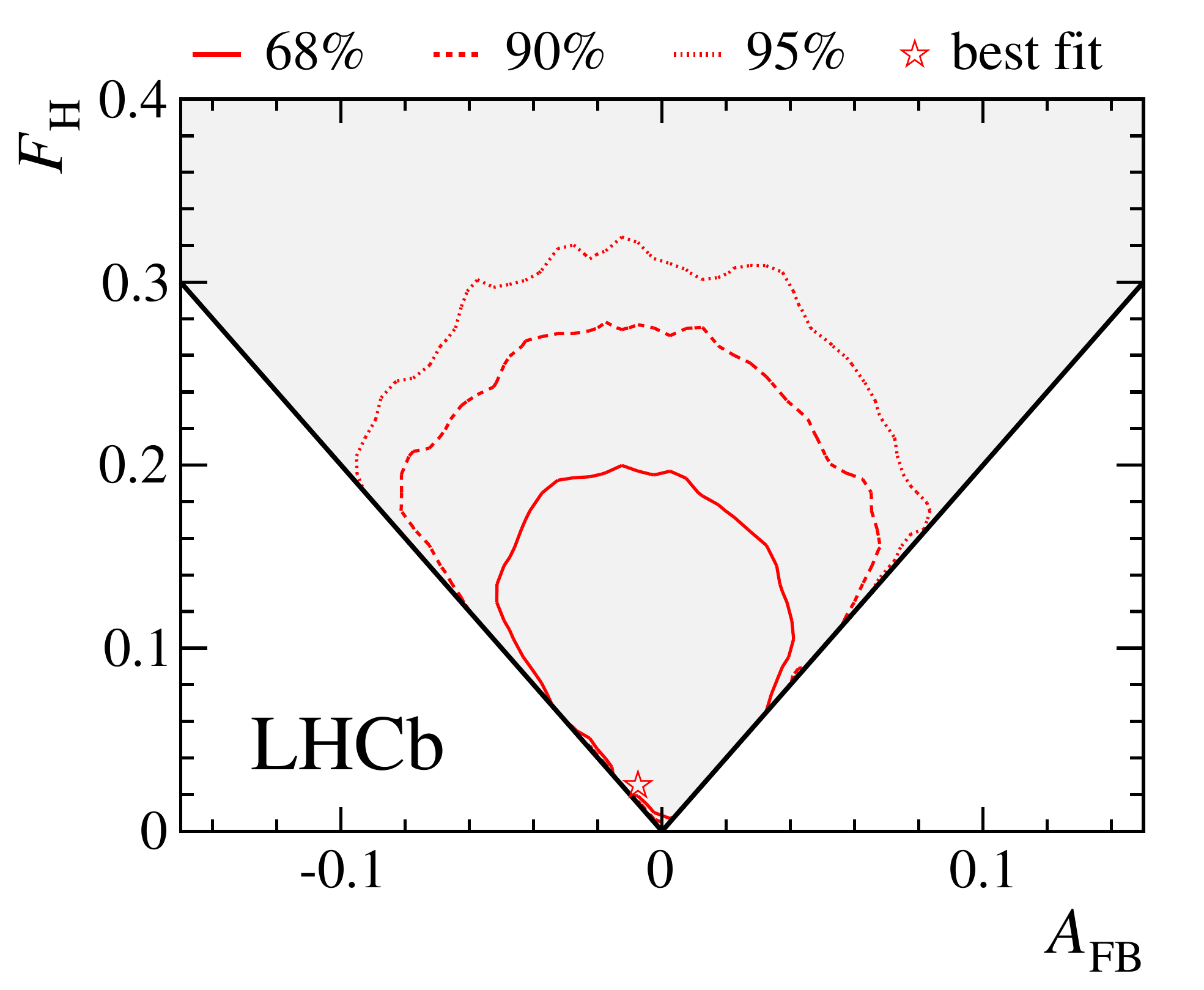}} \\
\subfigure[$6.00 < \qsq < 7.00\gev^{2}/c^{4}$]{\includegraphics[width=0.48\linewidth]{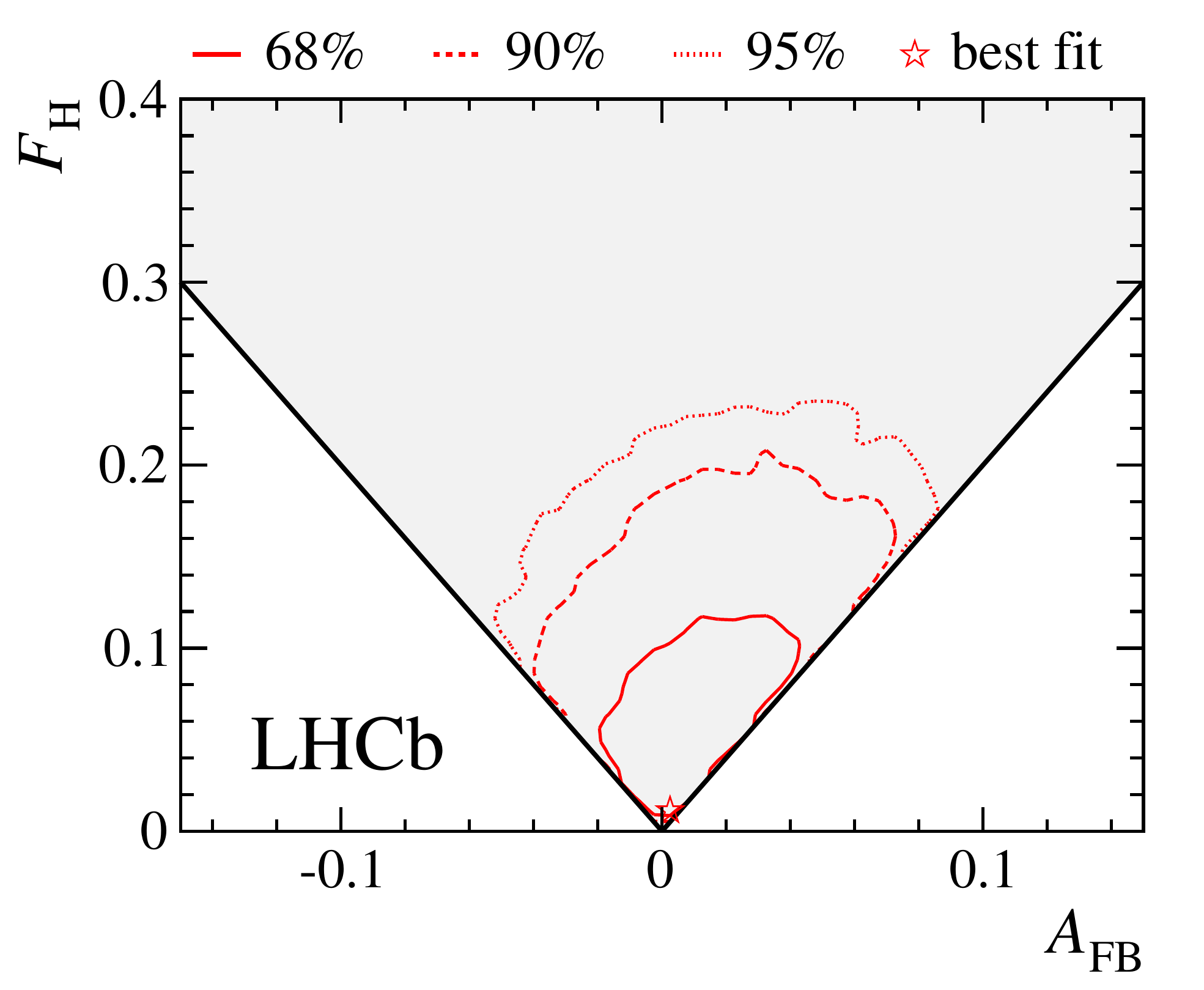}}
\subfigure[$7.00 < \qsq < 8.00\gev^{2}/c^{4}$]{\includegraphics[width=0.48\linewidth]{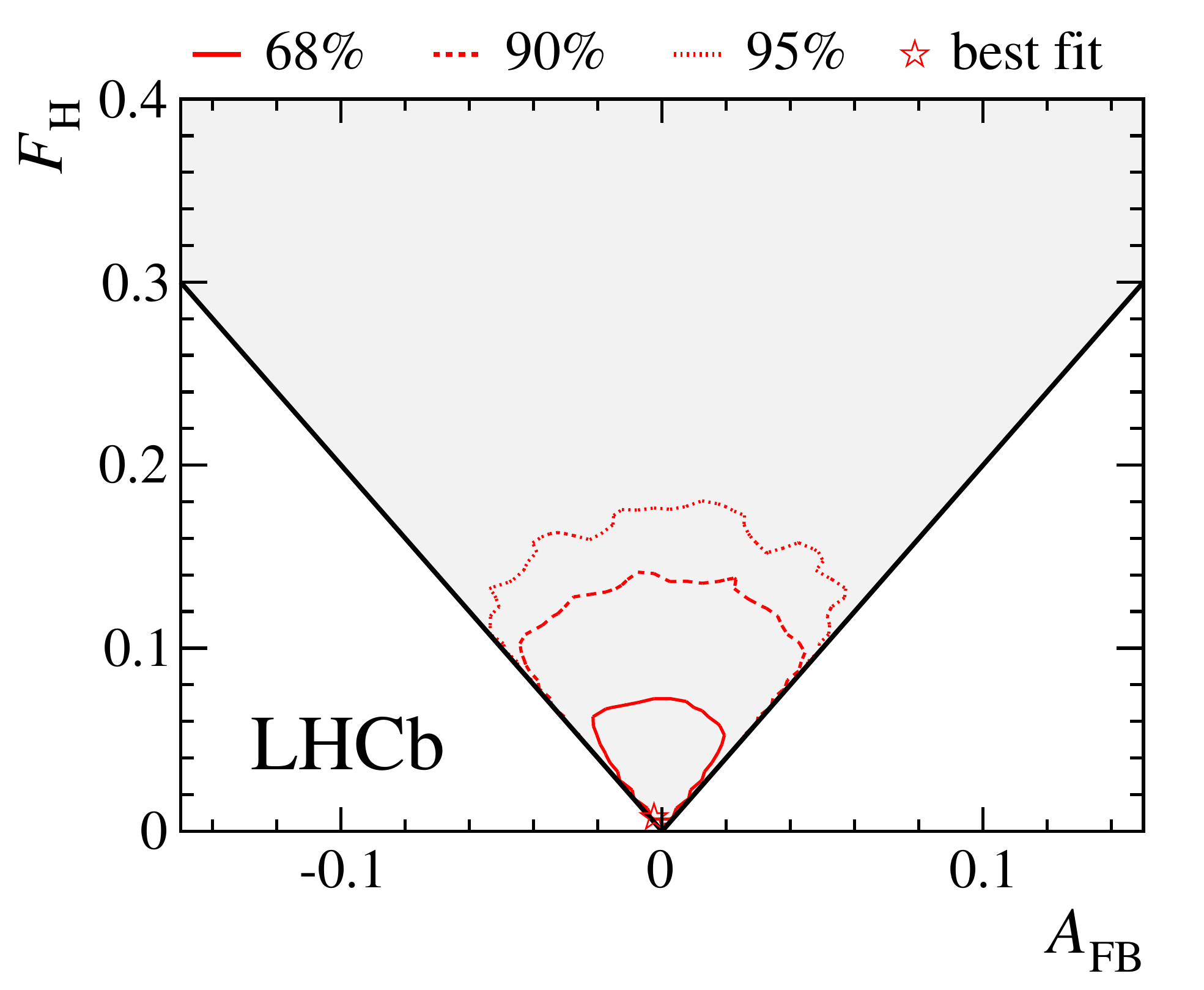}} \\
\caption{Two-dimensional confidence regions for $A_{\rm FB}$ and $F_{\rm H}$ for the decay \decay{\Bp}{\Kp\mumu} in the \qsq ranges (a) $4.00 < \qsq < 5.00\gevgevcccc$, (b) $5.00 < \qsq < 6.00\gevgevcccc$, (c) $6.00 < \qsq < 7.00\gevgevcccc$ and (d) $7.00 < \qsq < 8.00\gevgevcccc$. The confidence intervals are determined using the Feldman-Cousins technique and are purely statistical. The shaded (triangular) region illustrates the range of $A_{\rm FB}$ and $F_{\rm H}$ over which the signal angular distribution remains positive in all regions of phase-space.\label{fig:appendix:2D:B}}
\end{figure}

\begin{figure}[!h] 
\centering
\subfigure[$11.00 < \qsq < 11.75.00\gev^{2}/c^{4}$]{\includegraphics[width=0.48\linewidth]{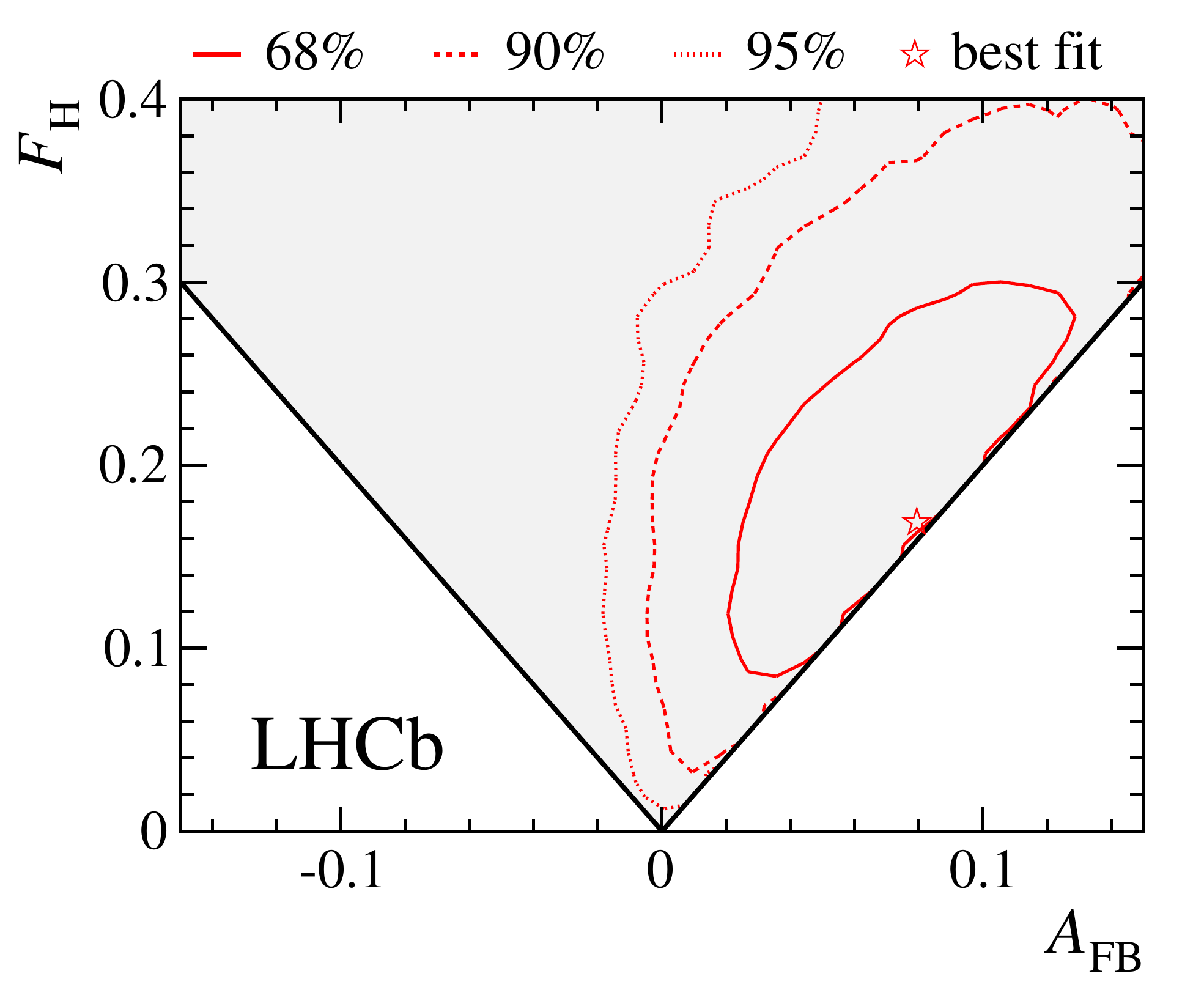}}
\subfigure[$11.75 < \qsq < 12.50\gev^{2}/c^{4}$]{\includegraphics[width=0.48\linewidth]{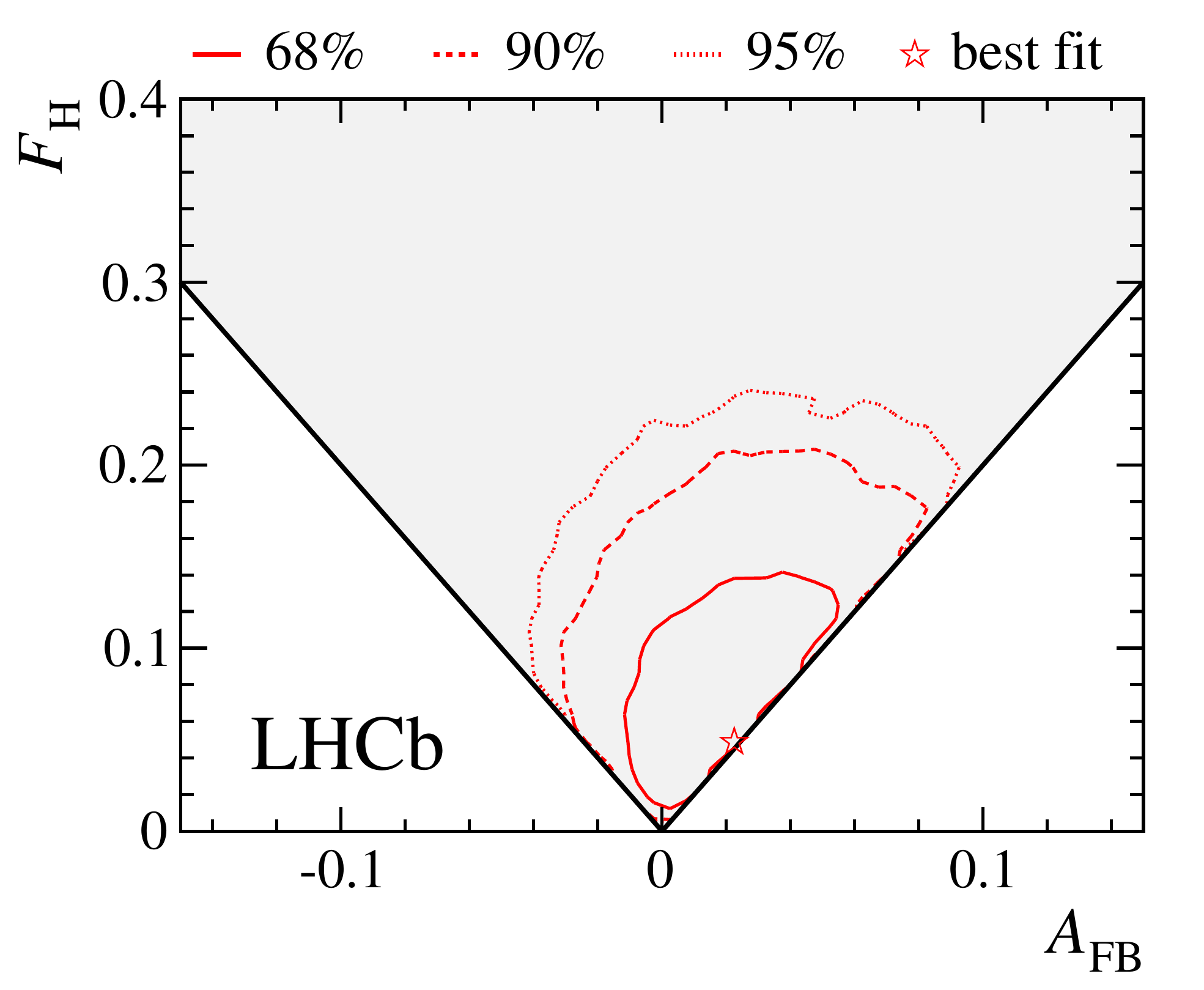}} \\
\subfigure[$15.00 < \qsq < 16.00\gev^{2}/c^{4}$]{\includegraphics[width=0.48\linewidth]{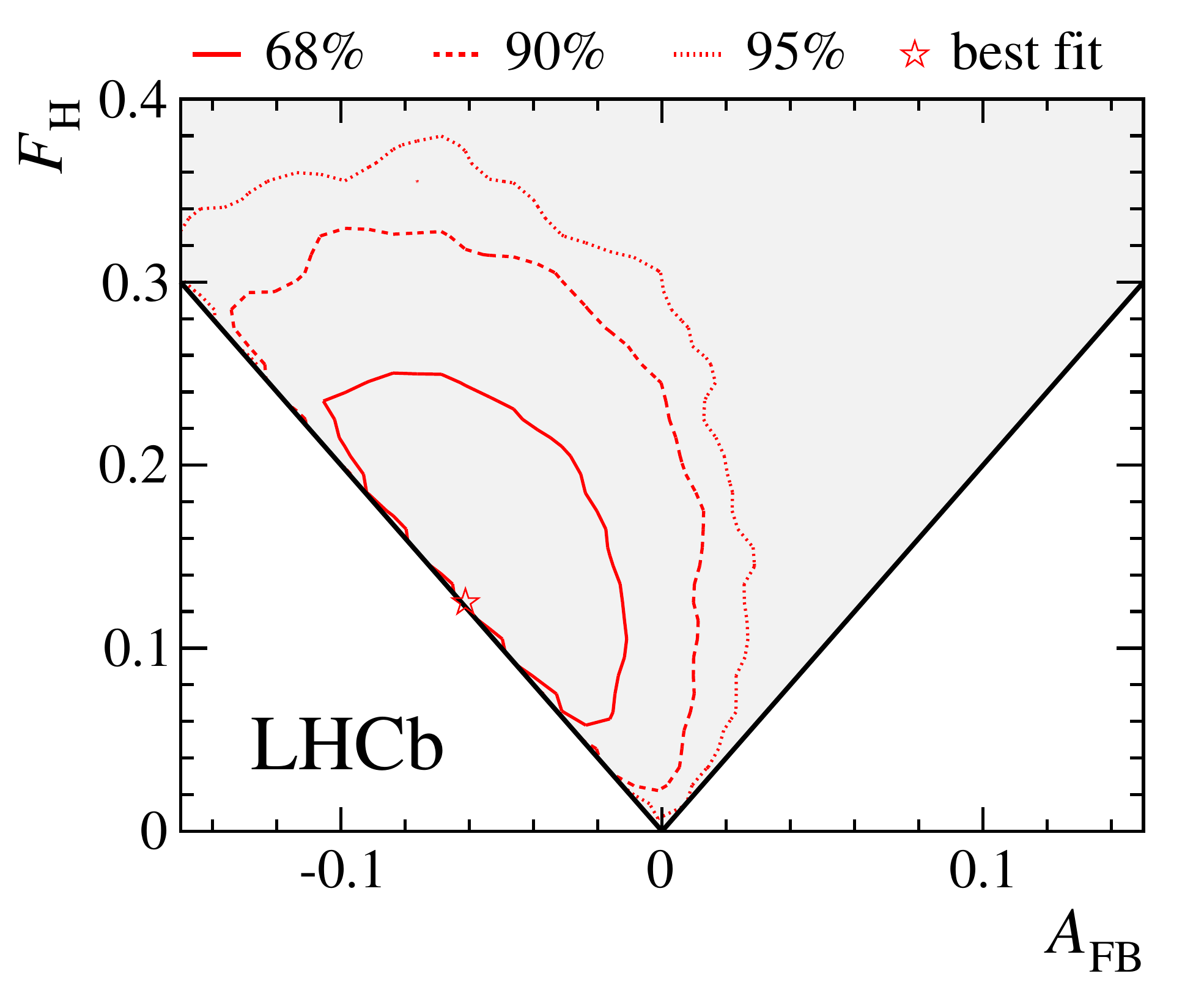}}

\caption{Two-dimensional confidence regions for $A_{\rm FB}$ and $F_{\rm H}$ for the decay \decay{\Bp}{\Kp\mumu} in the \qsq ranges (a) $11.00 < \qsq < 11.75\gevgevcccc$,  (b) $11.75 < \qsq < 12.50\gevgevcccc$ and (c) $15.00 < \qsq < 16.00\gevgevcccc$. The confidence intervals are determined using the Feldman-Cousins technique and are purely statistical. The shaded (triangular) region illustrates the range of $A_{\rm FB}$ and $F_{\rm H}$ over which the signal angular distribution remains positive in all regions of phase-space.\label{fig:appendix:2D:C}}
\end{figure}

\begin{figure}[!h] 
\centering
\subfigure[$16.00 < \qsq < 17.00\gev^{2}/c^{4}$]{\includegraphics[width=0.48\linewidth]{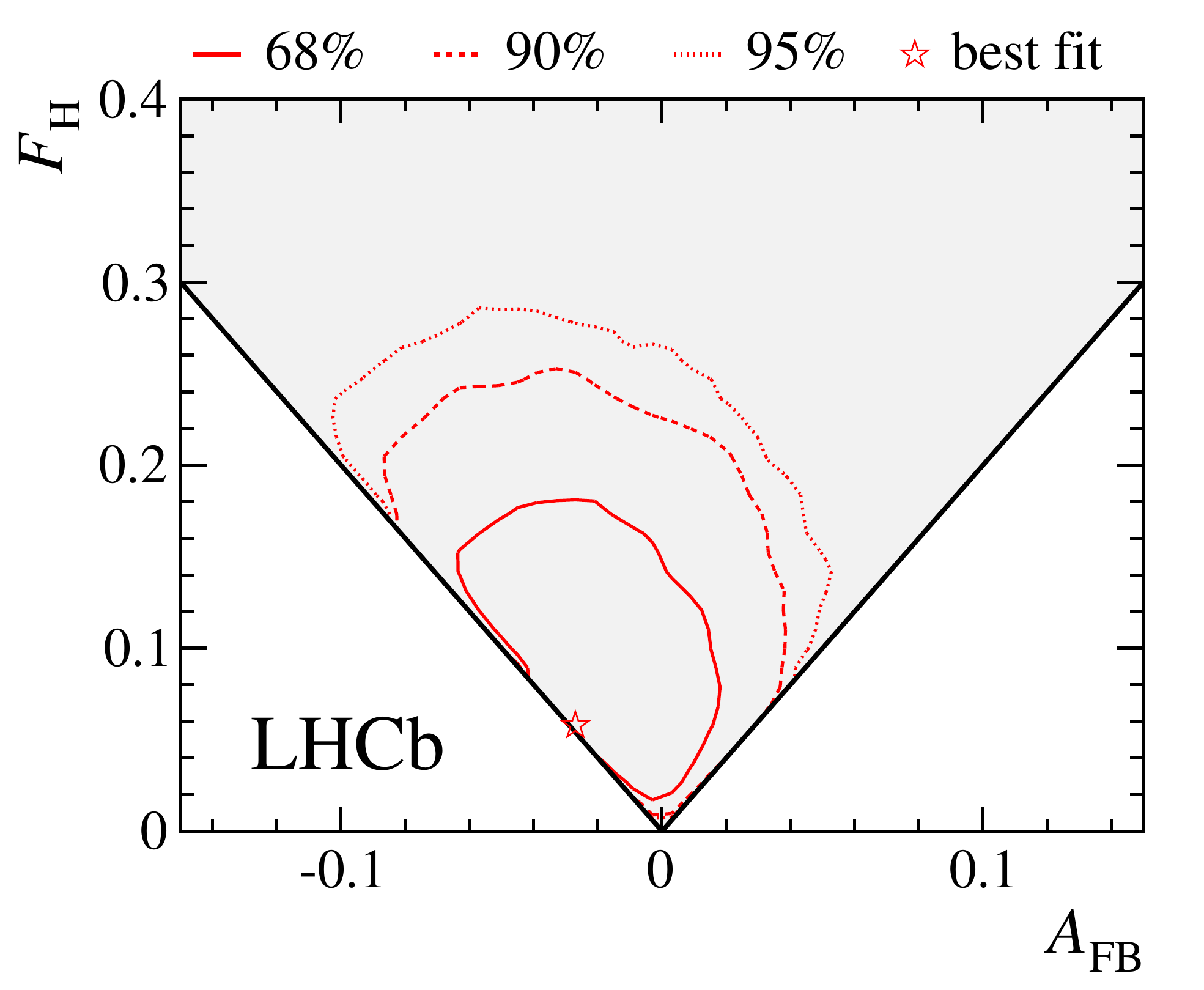}} 
\subfigure[$17.00 < \qsq < 18.00\gev^{2}/c^{4}$]{\includegraphics[width=0.48\linewidth]{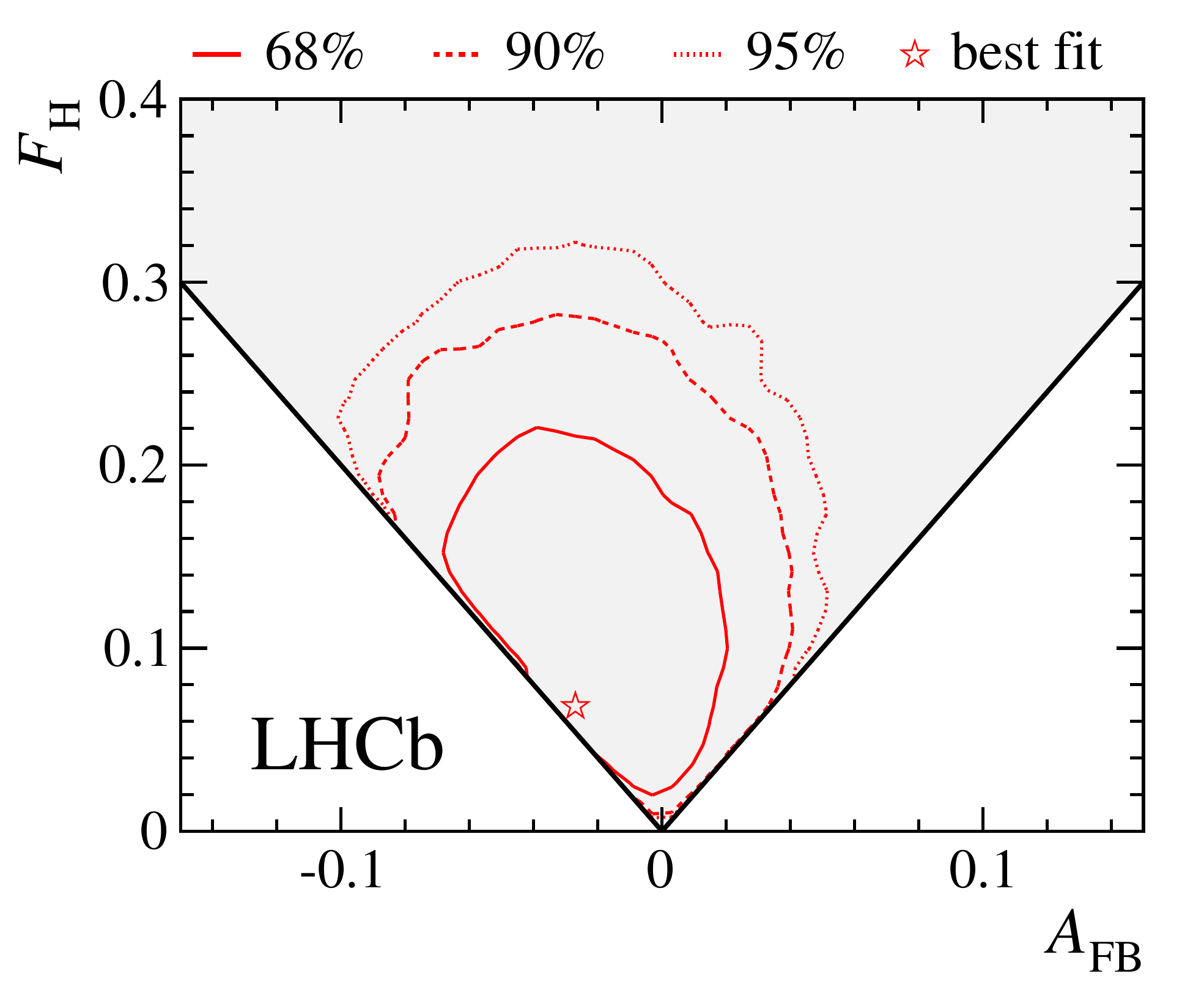}} \\ 
\subfigure[$18.00 < \qsq < 19.00\gev^{2}/c^{4}$]{\includegraphics[width=0.48\linewidth]{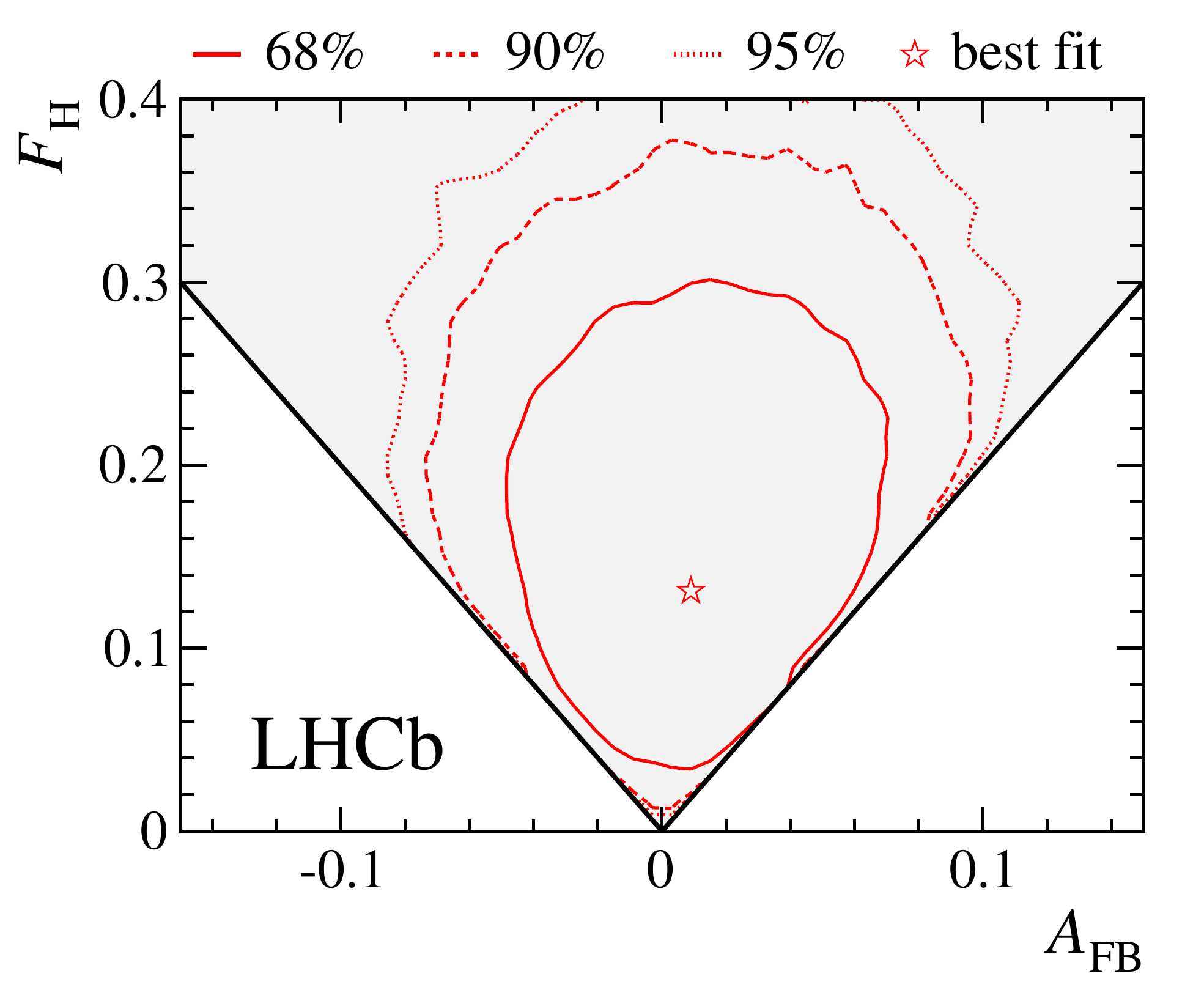}} 
\caption{Two-dimensional confidence regions for $A_{\rm FB}$ and $F_{\rm H}$ for the decay \decay{\Bp}{\Kp\mumu} in the \qsq ranges (a) $16.00 < \qsq < 17.00\gevgevcccc$, (b) $17.00 < \qsq < 18.00\gevgevcccc$ and (c) $18.00 < \qsq < 19.00\gevgevcccc$. The confidence intervals are determined using the Feldman-Cousins technique and are purely statistical. The shaded (triangular) region illustrates the range of $A_{\rm FB}$ and $F_{\rm H}$ over which the signal angular distribution remains positive in all regions of phase-space.\label{fig:appendix:2D:D}}
\end{figure}

\begin{figure}[!h] 
\centering
\subfigure[$19.00 < \qsq < 20.00\gev^{2}/c^{4}$]{\includegraphics[width=0.48\linewidth]{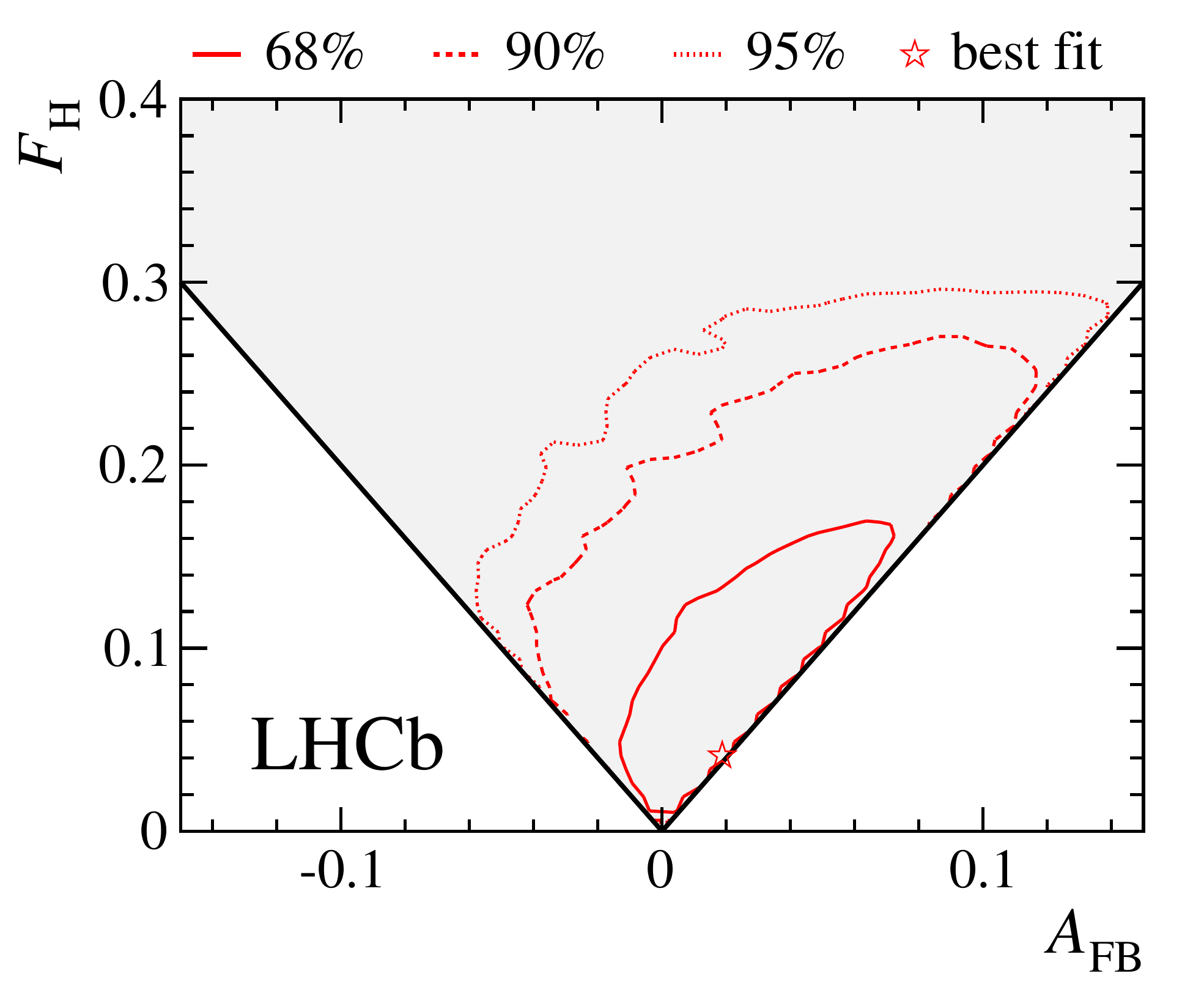}}
\subfigure[$20.00 < \qsq < 21.00\gev^{2}/c^{4}$]{\includegraphics[width=0.48\linewidth]{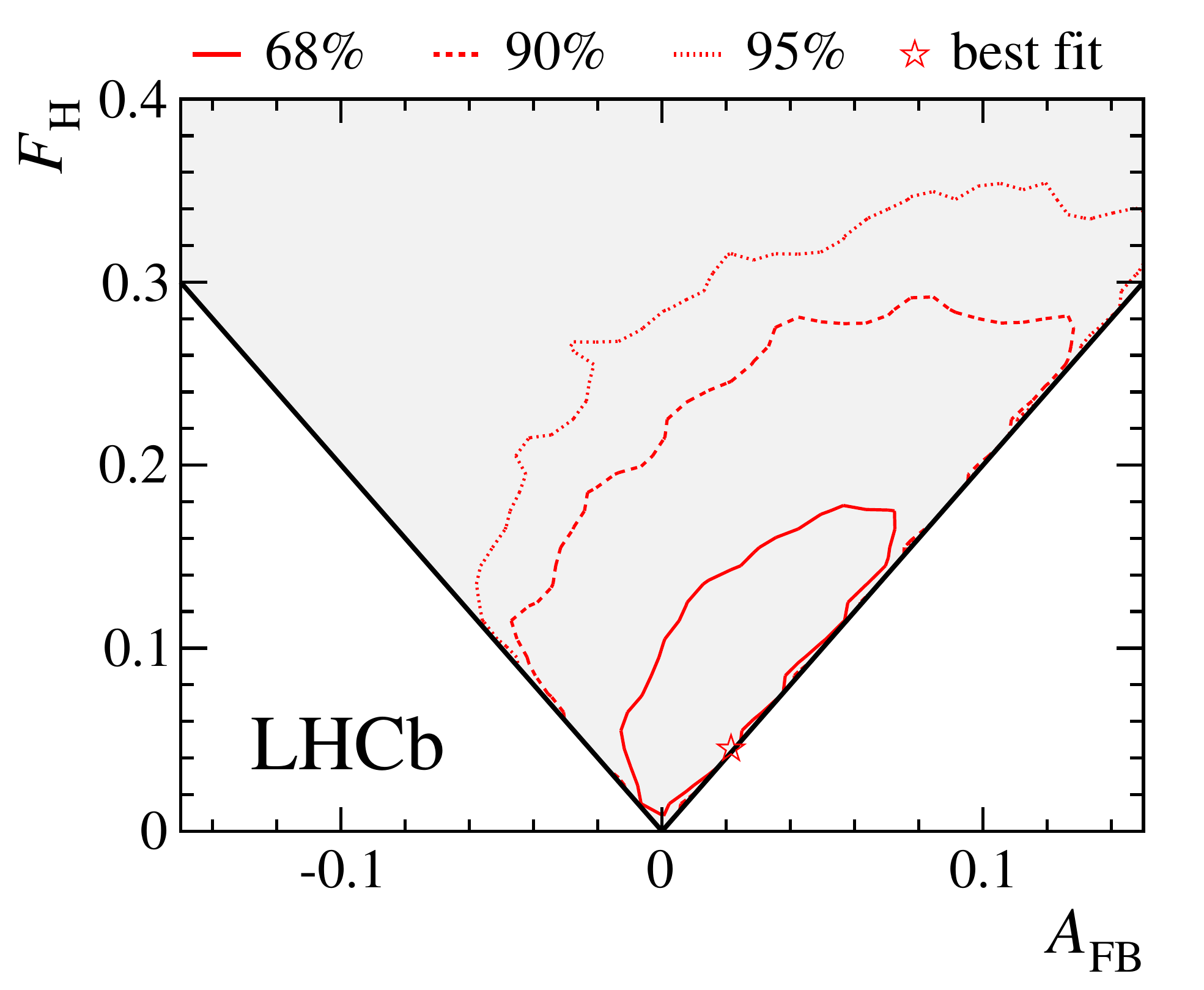}} \\
\subfigure[$21.00 < \qsq < 22.00\gev^{2}/c^{4}$]{\includegraphics[width=0.48\linewidth]{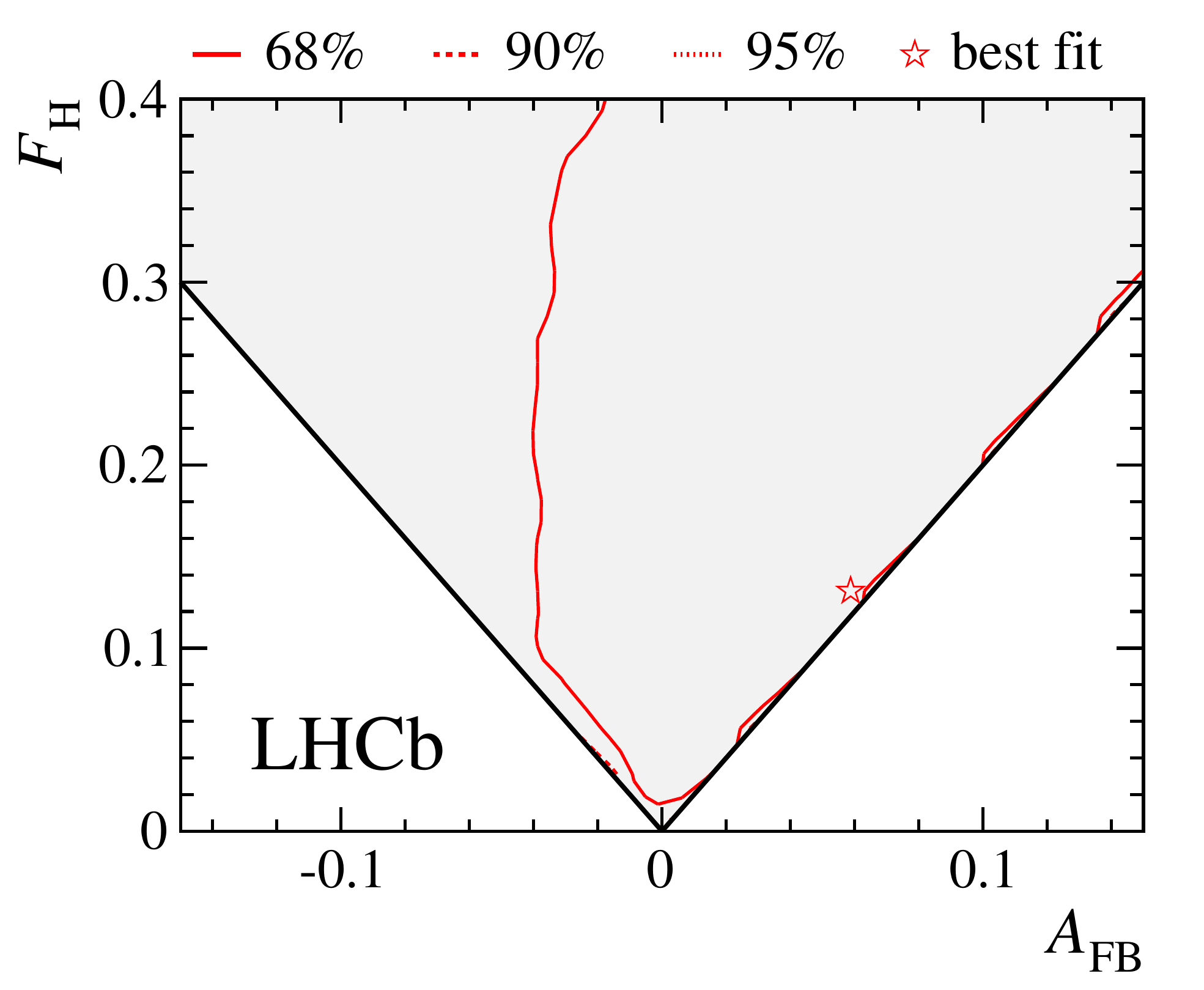}}
\caption{Two-dimensional confidence regions for $A_{\rm FB}$ and $F_{\rm H}$ for the decay \decay{\Bp}{\Kp\mumu} in the \qsq ranges (a) $19.00 < \qsq < 20.00\gevgevcccc$, (b) $20.00 < \qsq < 21.00\gevgevcccc$ and (c) $21.00 < \qsq < 22.00\gevgevcccc$. The confidence intervals are determined using the Feldman-Cousins technique and are purely statistical. The shaded (triangular) region illustrates the range of $A_{\rm FB}$ and $F_{\rm H}$ over which the signal angular distribution remains positive in all regions of phase-space.\label{fig:appendix:2D:E}}
\end{figure}


\clearpage

\end{document}